\documentclass[aps,prl,twocolumn,superscriptaddress,floatfix]{revtex4-2}
\usepackage{amsmath,amssymb,amsfonts,physics,bm}
\usepackage{graphicx}
\usepackage{microtype}
\usepackage{xurl}
\usepackage{hyperref}
\hypersetup{
  colorlinks=true,
  citecolor=blue,
  linkcolor=blue,
  urlcolor=blue,
  breaklinks=true
}
\usepackage{etoolbox}
\AtBeginEnvironment{thebibliography}{\small\sloppy\emergencystretch=8em\relax}
\usepackage{siunitx}
\usepackage{comment}
\usepackage[normalem]{ulem}
\usepackage{capt-of}
\usepackage{multirow}

\begin{document}
\title{Short Gravitational-Wave Transients as Probes of Cosmic Domain Walls}
\author{T\"{o}re Boybeyi}
\email{boybe001@umn.edu}
\affiliation{School of Physics \& Astronomy, University of Minnesota, Minneapolis, 55455, MN, USA}
\author{Do\u{g}a Veske}
\email{veske@metu.edu.tr}
\affiliation{Fizik B\"ol\"um\"u, Orta Do\u{g}u Teknik \"Universitesi, \c{C}ankaya, Ankara 06800, T\"urkiye}
\affiliation{Uzay ve H{\i}zland{\i}r{\i}c{\i} Teknolojileri Uygulama ve Ara\c{s}t{\i}rma Merkezi, Orta Do\u{g}u Teknik \"Universitesi, \c{C}ankaya, Ankara 06800, T\"urkiye}
\affiliation{Columbia Astrophysics Laboratory, Columbia University in the City of New York, New York, NY 10027, USA}
\author{David Maibach}
\email{maibach@thphys.uni-heidelberg.de}
\affiliation{Institute for Theoretical Physics, University of Heidelberg, Philosophenweg 16,
D-69120	Heidelberg,
Germany}
\begin{abstract}
\textcolor{black}{GW190521 and }GW231123 have been reported as short-duration gravitational-wave transients consistent with very massive binary black hole (BBH) coalescences whose inferred parameters, i.e., exceptionally high total masses and spin magnitudes, challenge standard isolated binary stellar evolution. We test a topological dark matter (TDM) interpretation invoking cosmic domain walls by fitting a physically motivated domain wall template to the LIGO Hanford and Livingston strain data. The BBH hypothesis is individually favored, with $\log_{10}\mathcal{B}_{\rm BBH/TDM}=12.2$ and $11.3$ for GW231123 and GW190521, respectively. However, these values are lower than those typically recovered from matched maximum a posteriori BBH waveforms injected into nearby noise segments. We further perform, for the first time, a joint fit in which domain wall signals from a single underlying scalar field are constrained simultaneously by both events. Although not favored over BBH signals, we find the two events are consistent with a common scalar field, with shared TDM parameters agreeing across independent noise realizations and sky locations. We further find that injected TDM transients are systematically recovered under the BBH hypothesis with large spin parameters, revealing a morphological degeneracy that could mask genuine domain wall signals. This analysis demonstrates that multi-event parameter consistency tests provide a new discriminant for domain wall dark matter searches in upcoming observing runs.
\end{abstract}
\maketitle

\textit{Introduction.}---The nature of dark matter remains one of the outstanding problems in modern physics. Its nature is investigated using a wide range of theories \cite{Feng:2010gw, Arbey:2021gdg, Khoury:2015pea, Adams:2022pbo}, potentially featuring \textit{exotic} types of particles. While particle-like candidates suggest an investigation of dark matter's nature by means of particle-like interactions, the comparably small momenta of candidate particles render most interaction-based detection methods with particle physics experiments impractical. Instead, one can resort to an unconventional method based on instruments originally designed for gravitational wave detection: In recent proposals, the possibility of precision measurements of dark matter coupled to the gauge sector of the Standard Model using ground-based large-scale interferometers is highlighted \cite{Jaeckel:2016jlh, Grote:2019uvn,  Khoze:2021uim, Pospelov:2012mt, vermeulen2021direct}(also see \cite{Dai:2021boq}). Here, the detection strategy relies on the assumption that a sufficiently large accumulation of (very low mass) dark matter passes through a detector beam, inducing a temporal variation of fundamental physical ``constants'' via non-gravitational interactions between dark matter and the Standard Model sector.

The stable, nontrivial accumulation of dark matter field values can be achieved in various ways. The most plausible and well-studied scenarios for such configurations are, for instance, extended topological defects such as domain walls, a form of topological dark matter (TDM), which can arise when discrete symmetries break in the early Universe~\cite{Kibble:1976sj,Vilenkin:2000jqa,Vachaspati:2006zz}. \textcolor{black}{We use the term TDM in the sense of Refs.~\cite{Stadnik:2014cea,Derevianko:2013oaa,Roberts:2017hla,Heisenberg:2023urf}, for a topological defect of a dark-sector scalar; as shown in the Discussion, the walls considered here are necessarily an energetically subdominant component of the dark matter. }As a concrete example, a sine-Gordon-type scalar theory with a periodic potential admits kink solutions that interpolate between adjacent vacua, yielding a localized region of trapped energy identified with a domain wall \cite{rajaraman1982solitons,Manton:2004tk}. The wall thickness is set by the inverse of the small oscillation mass \textcolor{black}{($m_\phi$)} around a vacuum, while the wall tension (or trapped energy) depends on the symmetry-breaking scale and model parameters~\cite{Vilenkin:2000jqa,Vachaspati:2006zz}.

If a domain wall passes through the Earth, parity-even, non-gravitational couplings to Standard Model fields (e.g., operators such as $\phi^2 F_{\mu\nu}F^{\mu\nu}$ and $\phi^2 m_f \bar{\psi}_f \psi_f$, where $\phi$ denotes the dark matter field) can lead to transient changes in optical properties and in the detector response of laser interferometers~\cite{Stadnik:2014cea,Derevianko:2013oaa,Roberts:2017hla,Grote:2019uvn}. Existing constraints still allow \textcolor{black}{up to one per day} domain wall passages through terrestrial detectors in some regions of parameter space, which motivates targeted searches for such transient signatures in interferometer data 
\cite{Derevianko:2013oaa,Roberts:2017hla,Grote:2019uvn,Heisenberg:2023urf}. A subset of short GW detections has therefore been examined for possible TDM interpretations in~\cite{Heisenberg:2023urf}. Notably, several works have investigated GW data in the context of dark matter \cite{vermeulen2021direct, Pierce:2018xmy,Guo:2019ker,LIGOScientific:2021ffg,theligoscientificcollaboration2025directmultimodeldarkmattersearch}, some targeting the gravitational-wave emission from the dark matter candidate itself \cite{Fell:2023mtf,Bhattacharya:2023stq,Dasgupta:2020mqg,Baryakhtar:2017ngi,Maselli:2020zgv,Maselli:2021men,Bartolo:2018evs}.

In case of the TDM, the effect in the interferometer output is triggered through different response channels. For example, an effective photon mass term can modify light propagation and produce a dispersive phase shift, while a kinetic type coupling can act through changes in electromagnetic parameters and associated effects on the optics~\cite{Heisenberg:2023urf}. Further, in a detector network, a single planar wall implies correlated signals across sites, with arrival times and relative amplitudes linked by the wall trajectory and orientation; these relations provide consistency checks that differ from those used for standard GW signals~\cite{Heisenberg:2023urf}.

Recently, an event with a short, few-cycle morphology similar to the transient shapes discussed in~\cite{Heisenberg:2023urf} has been reported as GW231123 \cite{LIGOScientific:2025rsn}.
It is a short transient observed in both LIGO interferometers~\cite{LIGOScientific:2014pky} during their fourth observing run, with only a few visible cycles and with reported parameters consistent with a high-mass, highly spinning binary black hole (BBH) merger (total mass $\sim 200$--$250\,M_\odot$, both components' spin parameter $\gtrsim0.8$)~\cite{LIGOScientific:2025rsn}. Such massive and highly spinning systems are difficult to produce through isolated binary stellar evolution and motivate tests of both hierarchical-merger scenarios and nonstandard transient models. One natural explanation invokes hierarchical mergers. However, the progenitor system would need more than one previous merger to achieve such masses~\cite{Li:2025pyo}. Nevertheless, the properties of GW231123 are not expected from typical star clusters~\cite{Passenger:2025acb}, but may be explained by nuclear star clusters~\cite{Paiella:2025qld}. These extraordinary properties motivate alternative explanations. GW231123 has been tested against exotic GW sources, such as cosmic strings~\cite{Cuceu:2025fzi,LIGOScientific:2025rsn}, where BBH waveforms were found to be favored. \textcolor{black}{The GW190521 event~\cite{LIGOScientific:2020iuh} from the third observing run also presents analogous challenges: its component masses fall within the pair-instability supernova mass gap, where no compact remnant is expected from standard stellar evolution~\cite{LIGOScientific:2020iuh}, making both events genuinely anomalous regardless of their physical interpretation}. We therefore carry out a dedicated test of TDM-induced interferometer transients in the regime where the resulting time-domain morphology can resemble very short BBH signals, following the signal modeling framework of Ref.~\cite{Heisenberg:2023urf}. A key feature of the domain wall hypothesis is that the scalar mass $m_\phi$, the photon sector parameter $m_0$, and the coupling $g_{\rm DW}$ are properties of the underlying field and must be shared across all events. By combining GW231123 with GW190521, previously identified as the most favorable short-duration candidate in~\cite{Heisenberg:2023urf}, we perform the first multi-event domain wall consistency test.

\textit{The domain wall model.}---We assume the local dark matter density comprises a topological domain wall formed by a real scalar field $\phi$~\cite{Derevianko:2013oaa,Stadnik:2014cea,Roberts:2017hla}, governed by a Lagrangian density with a periodic potential
\begin{align}
    \mathcal{L} &= \frac{1}{2}\partial_\mu\phi\partial^\mu\phi - \frac{2 m_\phi^2 f^2}{N_\phi^2} \sin^2\left(\frac{N_\phi \phi}{2f}\right),
\label{eq:scalar_lagrangian}
\end{align}
where $f$ is the symmetry-breaking scale and $m_\phi$ determines the effective mass. The integer parameter $N_\phi$ defines the discrete periodicity of the potential. \textcolor{black}{Periodic potentials of this form are generic for a pseudo-Nambu--Goldstone boson of a global $U(1)$ broken at the scale $f$ and explicitly broken to $\mathbb{Z}_{N_\phi}$ by non-perturbative effects; instantons in a confining dark sector, with $m_\phi\sim\Lambda_{\rm dark}^2/f$ and $\Lambda_{\rm dark}$ the confinement scale, are one realization~\cite{Vachaspati:2006zz,rajaraman1982solitons}. We adopt the sine-Gordon form because it is the minimal potential whose $\mathbb{Z}_{N_\phi}$ vacuum structure allows the coupling $\mathcal{S}(\phi)$ in Eq.~\eqref{eq:interaction_lagrangian} to vanish in all vacua while oscillating inside the wall, which is what gives rise to the discrete parameter $N_{\rm ratio}$ below, and because it admits the exact kink solution Eq.~\eqref{eq:soliton_profile}. The scale $f$ cancels in the observable detector response (see Eq.~\eqref{eq:waveform_general}), so neither $f$ nor $\Lambda_{\rm dark}$ is constrained by this analysis. }A planar wall moving with velocity $\bm{v}_{\rm DW}$ and normal $\hat{\bm{n}}$ connects adjacent vacua via the soliton solution
\begin{equation}
    \phi(u) = \frac{4f}{N_\phi} \arctan(e^{m_\phi u})
\label{eq:soliton_profile}
\end{equation}
with $u(\bm{r},t) = \hat{\bm{n}}\cdot\bm{r} - v_{\rm DW}(t-t_0)$. \textcolor{black}{Throughout, Eq.~\eqref{eq:scalar_lagrangian} is used as the local effective description of the wall at the crossing. Its parameters are treated as constant over the $\sim10^{4}$~km and $\lesssim0.1$~s relevant to a terrestrial crossing, which holds for any environmental dependence varying on astrophysical scales; whether $m_\phi$ and $f$ are also constant on cosmological scales depends on the origin of the walls, to which we return in the Discussion. Since $f$ cancels in the detector response, the inferred $m_\phi$ is operationally the inverse local wall thickness, $\delta_w\simeq\hbar c/m_\phi$.}
\textcolor{black}{We consider the interactions with the electromagnetic tensor $F_{\mu\nu}$ and the potential $A_\mu$, parameterized by a shape function $\mathcal{S}(\phi)=\sin^2\left(N_A \phi/f\right)$~\cite{Khoze:2021uim}:
\begin{subequations}
    \begin{equation}
    \mathcal{L}_{\rm int} \supset \frac{1}{4} g_{\rm DW} \mathcal{S}(\phi) F_{\mu\nu}F^{\mu\nu},
    \end{equation}
        \begin{equation}
\mathcal{L}_{\rm mass} \supset \frac{1}{2}\, m_0^2 \,\mathcal{S}(\phi)\, A_\mu A^\mu .
\end{equation}
\label{eq:interaction_lagrangian}
\end{subequations}}
The first interaction induces a spacetime-dependent variation in the fine-structure constant
$\delta\alpha/\alpha \simeq g_{\rm DW}\mathcal{S} (\phi)$~\cite{Stadnik:2014cea,Derevianko:2013oaa,Uzan:2010pm},
leading to shifts in the physical size of the test masses and in the interferometer optical path lengths. The dominant contribution from this coupling for ground-based detectors arises from center-of-mass accelerations driven by spatial gradients of the effective coupling. The second interaction is a phenomenological dispersive contribution parameterized by an effective photon mass $m_0$~\cite{Heisenberg:2023urf}. The envelope entering the total phase response can be written as
\begin{equation}
    \mathcal{S}(t) = \sin^2\left( 4 N_{\rm ratio} \arctan\left[\exp\left(\frac{2(t - t_{0})}{\tau_w}\right)\right] \right),
\label{eq:waveform_general}
\end{equation}
where $\tau_w = 2/(m_\phi v_{\rm DW})$ characterizes the signal duration and $N_{\rm ratio}\equiv N_A/N_\phi$. For detectors separated by a baseline $\bm{d}$, the relative signal arrival times are governed by
\begin{equation}
    \Delta t = \frac{\bm{d} \cdot \hat{\bm{n}}}{v_{\rm DW}},
\label{eq:time_delay}
\end{equation}
coupling the observed time delays directly to the wall velocity and direction. The full derivation of all response channels is given in the Supplemental Material~\cite{supp}\nocite{Varma:2019csw,LIGOScientific:2021usb,KAGRA:2021vkt} and in~\cite{Heisenberg:2023urf}.

\textit{Analysis and results.}---We perform a Bayesian analysis using strain data from the LIGO Hanford (H1) and LIGO Livingston (L1) interferometers from the first part of the fourth observing run \cite{LIGOScientific:2025snk}, utilizing the \texttt{bilby} inference library~\cite{Ashton:2018jfp} with the nested sampler \texttt{dynesty}~\cite{Speagle2020Dynesty}. Details of the likelihood, prior choices, and the BBH comparison analysis are given in the Supplemental Material ~\cite{supp}.

The TDM posterior for GW231123 peaks at 
$\log_{10}(m_\phi/{\rm eV})= -12.09\pm 0.11$, 
$\log_{10}(m_0/{\rm eV})= -11.19\pm 0.05$, 
$v_{\rm DW}\approx 2.6\times10^{-2}c$, and 
$N_{\rm ratio}=3$. The coupling \(g_{\rm DW}\) is weakly constrained, with the posterior spanning roughly four decades and remaining largely prior dominated.

We find that the BBH hypothesis is favored over the TDM hypothesis by $\log_{10} \mathcal{B}_{\rm BBH/TDM}=12.2$ (Table~\ref{tab:bayes}). Figure~\ref{fig:bbh_vs_dw_fit} compares the best fit reconstructions for both events. The wall template can reproduce a short transient with a duration scale compatible with the observed few cycle morphology of GW231123, so the maximum likelihood fit can appear qualitatively reasonable in terms of the network signal-to-noise ratio (SNR), with $\rm SNR_{\rm BBH}=20.8$ and $\rm SNR_{\rm TDM}=20.1$. The $3\%$ difference in peak SNR demonstrates that signal amplitude alone cannot discriminate between the two hypotheses; the Bayes factor advantage for BBH arises from waveform morphology and prior volume, not signal strength. The evidence difference, however, reflects the lower maximum likelihood of the TDM fit combined with the different effective prior volumes of the two models.

\begin{widetext}
\begin{center}
  \includegraphics[width=0.95\textwidth]{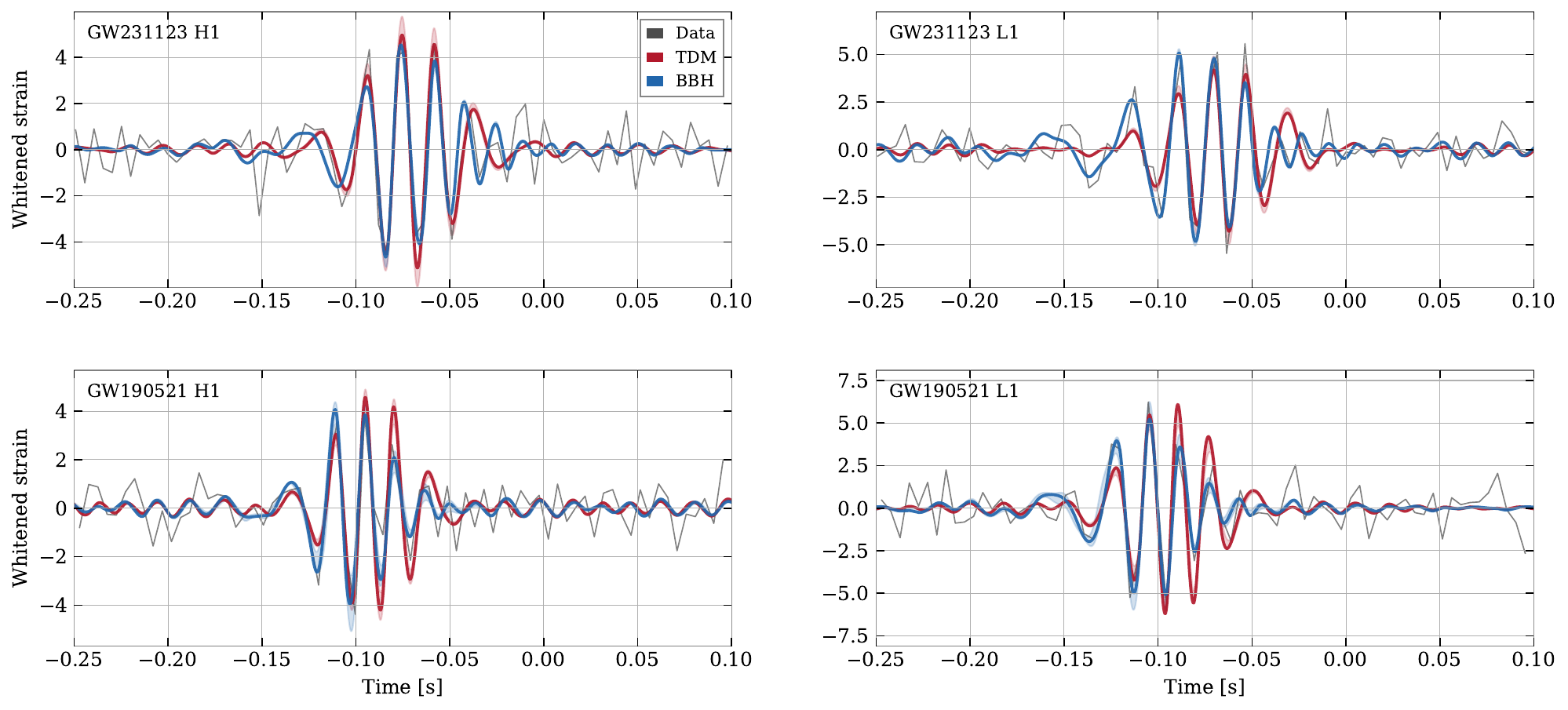}
  \captionof{figure}{\small Best fit reconstructions of the whitened strain for GW231123 (top) and GW190521 (bottom) under the BBH (blue) and TDM (red) hypotheses. Shaded regions show the 90\% credible interval from posterior draws. The TDM bands are obtained from the joint posterior with shared $m_\phi$, $m_0$, and $g_{\rm DW}$ across events.}
  \label{fig:bbh_vs_dw_fit}
\end{center}
\end{widetext}

 The smaller Occam penalty of the eight-parameter TDM model relative to the fifteen-parameter BBH model partially offsets the mismatch caused by the temporally symmetric envelope \(\mathcal{S}(t)\) in Eq.~\eqref{eq:waveform_general}. This symmetry follows from the constant velocity kink solution in Eq.~\eqref{eq:soliton_profile} and limits the ability of the TDM template to reproduce the asymmetric chirp structure of the data.

We next apply the same analysis to GW190521~\cite{LIGOScientific:2020iuh}, previously identified as the most favorable candidate for a TDM interpretation 
in~\cite{Heisenberg:2023urf}, obtaining $\log_{10}\mathcal{B}_{\rm BBH/TDM}=11.3$ with 
$\rm SNR_{\rm BBH}=15.2$ and $\rm SNR_{\rm TDM}=14.5$. 

\begin{table}[t]
\caption{Bayesian model comparison and injection calibration. We define \(\mathcal B\equiv\mathcal B_{\rm BBH/TDM}\). The quantity \(F_{\rm inj}\) is the fraction of matched MAP BBH injections with \(\log_{10}\mathcal B\leq\log_{10}\mathcal B_{\rm obs}\). The BBH injection medians are \(23.1\) for GW231123 and \(40.9\) for GW190521. The independent row sums the two per event log Bayes factors. The shared row is the joint fit with common \(m_\phi\), \(m_0\), and \(g_{\rm DW}\).}
\label{tab:bayes}
\begin{ruledtabular}
\begin{tabular}{lccc}
case & SNR & \(\log_{10}\mathcal B\) & \(F_{\rm inj}\) \\
GW231123 & \(20.8/20.1\) & \(12.2\) & \(1/50\) \\
GW190521 & \(15.2/14.5\) & \(11.3\) & \(4/50\) \\
independent & & \(23.5\) & \\
shared & & \(21.7\) & \\
\end{tabular}
\end{ruledtabular}
\end{table}

To calibrate this evidence ratio, we inject the maximum a posteriori BBH waveform for each event into 50 off-source segments drawn from the 192 s strain data surrounding the trigger and analyze the injections with the same conditioning and likelihood. The observed Bayes factors remain positive, favoring BBH over TDM, but their values are anomalously weak relative to matched BBH injections in comparable noise. For GW231123, \(\log_{10}\mathcal{B}_{\rm BBH/TDM}=12.2\), while the BBH injection median is \(23.1\) and \(F_{\rm inj}=1/50\); for GW190521, $\log_{10}\mathcal{B}_{\rm BBH/TDM}=11.3$ against a median of $40.9$ and $F_{\rm inj}=4/50$. The data therefore favor the BBH hypothesis, but not with the strength expected from genuine BBH signals. \textcolor{black}{We note that events identified by BBH template pipelines carry an inherent selection bias against the TDM hypothesis, as they are expected to resemble BBH waveforms by construction~\cite{Heisenberg:2023urf}; a dedicated unmodeled-burst or TDM-template search pipeline would provide a less biased test. Moreover, current pipelines form coincident candidates only within the window set by the GW propagation time at the speed of light ($\lesssim 10$~ms), whereas a wall whose normal $\hat{\bm n}$ is not aligned with the baseline produces a delay $\Delta t\sim |\bm d|/v_{\rm DW}$ of order seconds for $v_{\rm DW}\sim 10^{-3}$ to $10^{-2}c$, so a coherent transient with such a delay would be a decisive signature of a non gravitational wave origin but cannot be registered, leaving only the near coincident subset that is degenerate with BBH and is analyzed here.}

\begin{figure}[t]
  \centering
  \includegraphics[width=\columnwidth]{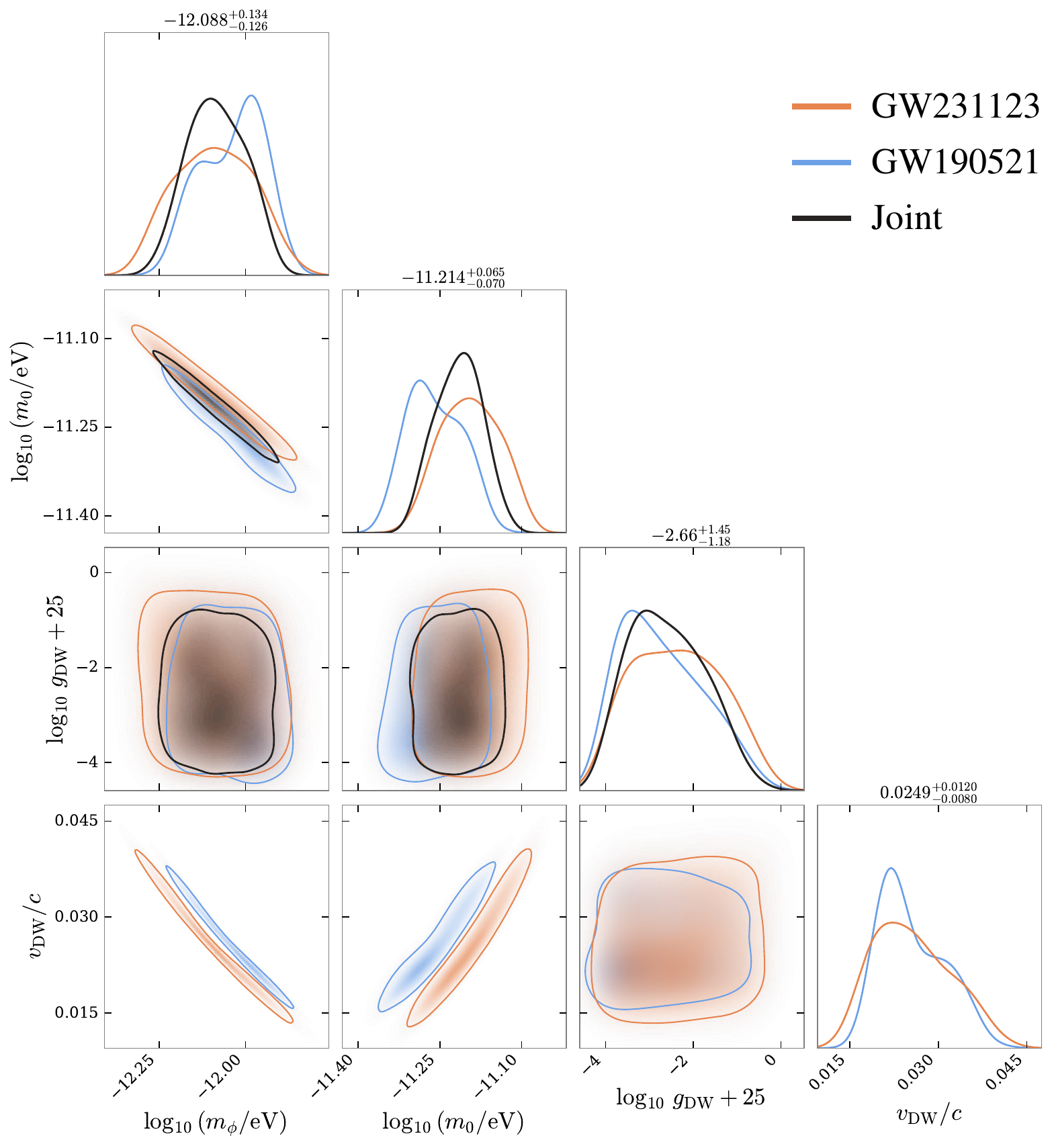}
  \caption{Posteriors on the shared dark matter parameters and wall velocity for GW231123 (orange) and GW190521 (blue). Black contours show the joint posterior on $m_\phi$, $m_0$, and $g_{\rm DW}$, obtained by joint fit. The wall velocity $v_{\rm DW}$ is event-specific; both individual posteriors are shown and no joint constraint on $v_{\rm DW}$ is imposed.}
  \label{fig:dw_corner2}
\end{figure}

Figure~\ref{fig:dw_corner2} compares the dark matter parameter 
posteriors. The two events agree to within $0.07$~dex ($0.4\sigma$) in $m_\phi$ and overlap in $m_0$, $v_{\rm DW}$, and $g_{\rm DW}$, and independently select the same $N_{\rm ratio}$, despite independent noise realizations and distinct sky locations. If the domain wall interpretation were correct, this consistency would reflect properties of a single underlying field rather than of individual events. The shared-parameter analysis is therefore a consistency test rather than a claim that either single event is TDM-favored. Events with inconsistent \(m_\phi\), \(m_0\), \(g_{\rm DW}\), or \(N_{\rm ratio}\) rapidly suppress the joint TDM evidence, while overlapping posteriors preserve the possibility of a common field origin. This makes the test increasingly sharp as the catalog of short-duration transients grows.

A separate cross-recovery test probes how a TDM transient is represented within the BBH hypothesis. We inject the GW231123 best fit TDM waveform into off-source noise and recover it under both hypotheses. The TDM recovery returns the injected domain wall parameter region, while the BBH recovery is pushed toward large spin magnitudes and a broad mass-distance region. This shows that spin and precession degrees of freedom can absorb short non-BBH morphology when only a few informative cycles are present.

\textit{Discussion.}---\textcolor{black}{The local template of Eqs.~\eqref{eq:scalar_lagrangian}--\eqref{eq:waveform_general} admits two distinct cosmological completions, which differ in what they imply for the constancy of the parameters and for the expected crossing rate. (a)~In a primordial scenario, $m_\phi$ and $f$ are constants fixed by microscopic physics and the walls form at a single epoch through the Kibble mechanism. A network of such walls in the scaling regime contains of order one wall per Hubble volume with curvature radius $\sim H^{-1}$~\cite{Vilenkin:2000jqa,Dankovsky:2024zvs}, so that the probability of a wall crossing the Earth within an observing run is $\sim v_{\rm DW}T_{\rm obs}H_0\sim10^{-12}$; a detectable rate therefore requires additional dynamics that leave a dense population of relic walls, as discussed below. (b)~Alternatively, a scalar field coupled non-minimally to matter can undergo a structure-induced phase transition at $z\lesssim1$~\cite{christiansen2026cosmicstringsdomainwalls, Hinterbichler_2010, Hinterbichler_2011}, with the discrete symmetry broken only on one side of a density threshold $\rho_{\rm crit}$ and walls forming at the boundaries between phases as cosmic structure evolves. In this case the parameters entering Eq.~\eqref{eq:scalar_lagrangian} are environment-dependent effective parameters, $m_\phi\to m_\phi(\rho)$ with $m_\phi^2(\rho)=m_\phi^2(1-\rho/\rho_{\rm crit})$ in the broken phase for a symmetron-type coupling, and what the fit measures is their value in the environment traversed by the wall. The number density and velocity distribution of the walls are then set by the late-time matter distribution rather than by scaling at a single epoch, and the relative velocity of the Earth through such a network is naturally of order $10^{-3}$ to $10^{-2}c$, set by the Milky Way bulk flow and substructure dynamics, consistent with the joint best-fit value. Neither completion is tested by the present analysis, which constrains only the local wall parameters; we note, however, that the consistency of $m_\phi v_{\rm DW}$, $m_0$, and $N_{\rm ratio}$ across two events is a necessary condition for both.}

Although the evidence favors the BBH interpretation, the TDM posterior parameters carry physical content worth noting in anticipation of future searches. The posterior median coupling $g_{\rm DW}\sim 10^{-27}$ corresponds to a fractional variation of the fine-structure constant $\delta\alpha/\alpha\sim 10^{-27}$; the posterior itself is broad (spanning the full prior), but even the median lies ten orders of magnitude below the sensitivity of current atomic clock searches for transient $\alpha$-variations~\cite{Roberts:2019sfo}, illustrating why an interferometric detection channel is necessary to probe this regime. Both events independently prefer $N_{\rm ratio}=3$, with GW231123 concentrating entirely at this value and GW190521 assigning $85\%$ posterior weight to it. Since $N_{\rm ratio}$ determines the number of oscillation cycles in the domain wall envelope, this agreement across two independent events with different detector noise realizations and sky locations is a necessary (though not sufficient) condition for a common underlying wall population. With five discrete values in the prior, chance agreement at the same $N_{\rm ratio}$ has prior probability $20\%$ per event; the observed joint selection across two independent events therefore constitutes a non-trivial cross-check.

\textcolor{black}{The preferred wall speed $v_{\rm DW}\sim 2.6\times10^{-2}c$ is driven by two 
requirements: the signal duration $\tau_w\sim 2/(m_\phi v_{\rm DW})$ must fall 
in the LIGO sensitivity band, and the intersite delay must fit within the analysis window. For the LIGO baseline $|\bm d|\approx 3000$~km, coherent fitting requires $v_{\rm DW}\gtrsim 10^{-3}c$. \textcolor{black}{In the primordial completion~(a), }the wall speed can be related to the formation epoch through the Hubble damping relation $\gamma v\,a^3=\mathrm{const}$~\cite{Avelino:2014xda}. Network 
simulations in the scaling regime yield relativistic velocities, giving $z_{\rm PT}\gtrsim 4$, with higher values for walls that formed above 
the network average speed. }

\textcolor{black}{The posteriors also reveal comparable magnitudes for the scalar mass and the photon sector parameter, with $\log_{10}(m_\phi/{\rm eV})\approx -12.1$ and $\log_{10}(m_0/{\rm eV})\approx -11.2$. These two parameters enter the signal through distinct channels: $m_\phi$ sets the wall thickness and hence the transient duration, while $m_0$ governs the dispersive phase shift. Notably, the dispersive interaction $m_0^2\,\mathcal{S}(\phi)\,A_\mu A^\mu$ and the scalar potential $\propto m_\phi^2\sin^2(N_\phi\phi/2f)$ share the same functional dependence on the field profile. If both mass scales arise from a common ultraviolet completion, their proximity would be expected.}

For an event rate of \(\mathcal O(1)\) per year, the local interwall spacing is \(D_{\rm local}\sim\mathcal O(100)\) AU, corresponding to \(n_{\rm DW,local}=D_{\rm local}^{-3}\approx 10^{-42}\,{\rm m}^{-3}\). Assuming a cold-DM-like halo enhancement \(\eta=\rho_{\rm DM,local}/\rho_{\rm DM,cosmic}\approx 3\times10^5\) (an order-of-magnitude estimate based on a standard NFW profile for the Milky Way halo), this corresponds to a cosmic mean spacing \(D_{\rm cosmic}\sim\mathcal O(0.1)\) pc. For wall tension \(\sigma\), the condition \(\rho_{\rm wall,local}=\sigma/D_{\rm local}\leq \epsilon\,\rho_{\rm DM,local}\), with \(\epsilon\) a subdominant fractional contribution, gives \(\sigma\leq \epsilon\,10^{15}\,{\rm eV}^3\). Thus, for \(\epsilon\sim\mathcal O(10^{-3})\), the model lies well within the Zel'dovich bound \cite{Zeldovich:1974uw} and modern constraints derived from it \cite{Saikawa_2017}. \textcolor{black}{Because $f$ cancels in the detector response, this bound is most usefully expressed as a bound on the symmetry-breaking scale: with $\sigma=8f^{2}m_\phi/N_\phi^{2}$~\cite{Heisenberg:2023urf} and the best-fit $m_\phi$, it requires $f\lesssim4\times10^{2}\,{\rm GeV}\,N_\phi\,(\epsilon/10^{-3})^{1/2}$. The same bound controls the energy in the homogeneous mode of $\phi$. In the primordial scenario the field starts oscillating at $T_{\rm osc}\sim(m_\phi M_{\rm Pl})^{1/2}\approx50$~MeV and its energy density today, $\sim m_\phi^{2}f^{2}(T_0/T_{\rm osc})^{3}\lesssim10^{-34}\,{\rm eV^{4}}$, is negligible; overproduction of dark matter by misalignment would require $f\gtrsim10^{14}$~GeV, far outside the allowed window. In the structure-induced scenario the energy released when the field settles into the broken phase at $z\lesssim1$, $\Delta V\sim2m_\phi^{2}f^{2}/N_\phi^{2}$, is not diluted afterwards and must itself be subdominant, which requires \textcolor{black}{$f\lesssim10^{-2}$~GeV}; in the symmetron realization~\cite{Hinterbichler_2010, Hinterbichler_2011} this is automatic, $\Delta V\sim\rho_{\rm crit}(M/M_{\rm Pl})^{2}\lesssim10^{-6}\rho_{\rm crit}$ for a coupling scale $M\lesssim10^{-3}M_{\rm Pl}$. In either case the scalar sector, walls and homogeneous mode alike, is energetically subdominant, and the detectability of a crossing rests on the couplings $g_{\rm DW}$ and $m_0$ rather than on the wall energy.}

The walls therefore constitute a subdominant dark matter component and do not violate existing observational constraints. In particular, the scenario remains viable within the symmetron parameter window \cite{Hinterbichler_2010, Hinterbichler_2011} not yet excluded by atom interferometry \cite{Burrage_2016, Panda:2023nir} or neutron bouncing experiments \cite{Cronenberg_2018}. The TDM configuration considered here therefore does not require additional fine-tuning beyond that already present in symmetron dark energy models.

\textcolor{black}{We stress that a detectable crossing rate is not a generic expectation for a primordial network. In the scaling regime, essentially all of the wall area resides in a single long wall per Hubble volume with curvature radius $\sim H^{-1}$~\cite{Dankovsky:2024zvs}, and the probability that it sweeps through the Earth during an observing run is $\sim v_{\rm DW}T_{\rm obs}H_0\sim10^{-12}$. The exact periodicity of Eq.~\eqref{eq:scalar_lagrangian} moreover yields topologically stable walls whose scaling network would overclose the universe, so that a viable primordial scenario requires either a small explicit bias destabilizing the network or dilution by inflation, as for QCD axion domain walls with $N_{\rm DW}>1$~\cite{Vilenkin:2000jqa,Vachaspati:2006zz}. An $\mathcal O(1)\,{\rm yr}^{-1}$ rate therefore presupposes a dense local population of relic walls with mean spacing $\lesssim10^{2}$~AU (see above), which must originate either from a primordial network that did not reach scaling or from the late-time, structure-induced mechanism of Refs.~\cite{christiansen2026cosmicstringsdomainwalls, Hinterbichler_2010, Hinterbichler_2011}, in which the walls are anchored to the matter distribution and their separation is set by the structure scale rather than by the horizon. The existence of such a population is the central theoretical requirement of the scenario tested here; the analysis itself only requires the wall to be planar across the detector network (Supplemental Material \cite{supp}).}

\textit{Conclusion.}---Exceptionally heavy candidates such as GW231123 \textcolor{black}{and GW190521 }contain only a few visible cycles, motivating explicit hypothesis tests on the strain data where limited cycle content can impede qualitative discrimination between competing models.

We performed such a test for a domain wall interpretation, treating the domain wall as a macroscopic field configuration that can induce a transient phase response during a crossing. In the model adopted here, the duration is set by the wall thickness and speed, and the relative timing between sites is fixed by the wall trajectory. We fit this domain wall template to the Hanford and Livingston data and carried out Bayesian model comparison against a BBH waveform model using the same conditioned data segment for both hypotheses. The evidence favors the BBH hypothesis for GW231123 \textcolor{black}{and GW190521} in our baseline analysis. The injection study does not overturn this preference, but shows that the observed Bayes factor is anomalously weak relative to matched maximum a posteriori BBH injections in nearby noise, signaling an atypical signal morphology that the BBH model only marginally accommodates.

Short-duration transients beyond the BBH class remain an important target because they can encode rich physics, including nonstandard compact object dynamics and macroscopic field or defect scenarios. The expected rate of crossings is model-dependent, scaling with wall number density, velocity, and the detector cross section\textcolor{black}{, and, as discussed above, presupposes a dense local wall population that a primordial scaling network cannot provide}.
Conversely, the temporal symmetry of the domain wall template provides a qualitative discriminant: a future short transient with a manifestly symmetric time-frequency structure would be difficult to accommodate as a BBH but natural for a domain wall crossing.

In this context, explicit domain wall templates provide a concrete way to test one such possibility directly on the strain data. Our results support the standard BBH interpretation of GW231123 \textcolor{black}{and GW190521} within the domain wall framework adopted here. At the same time, the consistency of the recovered dark matter parameters across two independent events establishes a baseline for multi-event domain wall searches. As the catalog of short-duration transients grows in upcoming observing runs, the joint analysis framework demonstrated here can accumulate evidence for or against a shared dark matter field origin with increasing statistical power.

\section*{Acknowledgments}
We thank H. Guo for useful comments. This document has been assigned to the LVC Document number: P2600056. DV is grateful to The Scientific and Technological Research Council of T\"urkiye for their support through the grants 123C484 and 125F379. DV acknowledges further support from the European Union’s Horizon Europe Research and Innovation Programme under the Marie Skłodowska-Curie
Actions (MSCA) COFUND Programme with the grant
number 101081645 and The Scientific and Technological Research Council of T\"urkiye with the grant number 123C213. TB was in part supported by the NSF grant PHY-2409173 and Aneesur Rahman Award. This research has made use of data or software obtained from the Gravitational Wave Open Science Center (gwosc.org), a service of the LIGO Scientific Collaboration, the Virgo Collaboration, and KAGRA. This material is based upon work supported by LIGO Laboratory which is a major facility fully funded by the National Science Foundation, as well as the Science and Technology Facilities Council (STFC) of the United Kingdom, the Max-Planck-Society (MPS), and the State of Niedersachsen/Germany for support of the construction of Advanced LIGO and construction and operation of the GEO600 detector. Additional support for Advanced LIGO was provided by the Australian Research Council. Virgo is funded, through the European Gravitational Observatory (EGO), by the French Centre National de Recherche Scientifique (CNRS), the Italian Istituto Nazionale di Fisica Nucleare (INFN) and the Dutch Nikhef, with contributions by institutions from Belgium, Germany, Greece, Hungary, Ireland, Japan, Monaco, Poland, Portugal, Spain. KAGRA is supported by Ministry of Education, Culture, Sports, Science and Technology (MEXT), Japan Society for the Promotion of Science (JSPS) in Japan; National Research Foundation (NRF) and Ministry of Science and ICT (MSIT) in Korea; Academia Sinica (AS) and National Science and Technology Council (NSTC) in Taiwan.


\begin{thebibliography}{56}%
\makeatletter
\providecommand \@ifxundefined [1]{%
 \@ifx{#1\undefined}
}%
\providecommand \@ifnum [1]{%
 \ifnum #1\expandafter \@firstoftwo
 \else \expandafter \@secondoftwo
 \fi
}%
\providecommand \@ifx [1]{%
 \ifx #1\expandafter \@firstoftwo
 \else \expandafter \@secondoftwo
 \fi
}%
\providecommand \natexlab [1]{#1}%
\providecommand \enquote  [1]{``#1''}%
\providecommand \bibnamefont  [1]{#1}%
\providecommand \bibfnamefont [1]{#1}%
\providecommand \citenamefont [1]{#1}%
\providecommand \href@noop [0]{\@secondoftwo}%
\providecommand \href [0]{\begingroup \@sanitize@url \@href}%
\providecommand \@href[1]{\@@startlink{#1}\@@href}%
\providecommand \@@href[1]{\endgroup#1\@@endlink}%
\providecommand \@sanitize@url [0]{\catcode `\\12\catcode `\$12\catcode `\&12\catcode `\#12\catcode `\^12\catcode `\_12\catcode `\%12\relax}%
\providecommand \@@startlink[1]{}%
\providecommand \@@endlink[0]{}%
\providecommand \url  [0]{\begingroup\@sanitize@url \@url }%
\providecommand \@url [1]{\endgroup\@href {#1}{\urlprefix }}%
\providecommand \urlprefix  [0]{URL }%
\providecommand \Eprint [0]{\href }%
\providecommand \doibase [0]{https://doi.org/}%
\providecommand \selectlanguage [0]{\@gobble}%
\providecommand \bibinfo  [0]{\@secondoftwo}%
\providecommand \bibfield  [0]{\@secondoftwo}%
\providecommand \translation [1]{[#1]}%
\providecommand \BibitemOpen [0]{}%
\providecommand \bibitemStop [0]{}%
\providecommand \bibitemNoStop [0]{.\EOS\space}%
\providecommand \EOS [0]{\spacefactor3000\relax}%
\providecommand \BibitemShut  [1]{\csname bibitem#1\endcsname}%
\let\auto@bib@innerbib\@empty
\bibitem [{\citenamefont {Feng}(2010)}]{Feng:2010gw}%
  \BibitemOpen
  \bibfield  {author} {\bibinfo {author} {\bibfnamefont {J.~L.}\ \bibnamefont {Feng}},\ }\href {https://doi.org/10.1146/annurev-astro-082708-101659} {\bibfield  {journal} {\bibinfo  {journal} {Ann. Rev. Astron. Astrophys.}\ }\textbf {\bibinfo {volume} {48}},\ \bibinfo {pages} {495} (\bibinfo {year} {2010})},\ \Eprint {https://arxiv.org/abs/1003.0904} {arXiv:1003.0904 [astro-ph.CO]} \BibitemShut {NoStop}%
\bibitem [{\citenamefont {Arbey}\ and\ \citenamefont {Mahmoudi}(2021)}]{Arbey:2021gdg}%
  \BibitemOpen
  \bibfield  {author} {\bibinfo {author} {\bibfnamefont {A.}~\bibnamefont {Arbey}}\ and\ \bibinfo {author} {\bibfnamefont {F.}~\bibnamefont {Mahmoudi}},\ }\href {https://doi.org/10.1016/j.ppnp.2021.103865} {\bibfield  {journal} {\bibinfo  {journal} {Prog. Part. Nucl. Phys.}\ }\textbf {\bibinfo {volume} {119}},\ \bibinfo {pages} {103865} (\bibinfo {year} {2021})},\ \Eprint {https://arxiv.org/abs/2104.11488} {arXiv:2104.11488 [hep-ph]} \BibitemShut {NoStop}%
\bibitem [{\citenamefont {Khoury}(2015)}]{Khoury:2015pea}%
  \BibitemOpen
  \bibfield  {author} {\bibinfo {author} {\bibfnamefont {J.}~\bibnamefont {Khoury}},\ }in\ \href@noop {} {\emph {\bibinfo {booktitle} {{50th Rencontres de Moriond on Gravitation: 100 years after GR}}}}\ (\bibinfo {year} {2015})\ pp.\ \bibinfo {pages} {35--42},\ \Eprint {https://arxiv.org/abs/1507.03013} {arXiv:1507.03013 [astro-ph.CO]} \BibitemShut {NoStop}%
\bibitem [{\citenamefont {Adams}\ \emph {et~al.}(2022)\citenamefont {Adams} \emph {et~al.}}]{Adams:2022pbo}%
  \BibitemOpen
  \bibfield  {author} {\bibinfo {author} {\bibfnamefont {C.~B.}\ \bibnamefont {Adams}} \emph {et~al.},\ }in\ \href@noop {} {\emph {\bibinfo {booktitle} {{Snowmass 2021}}}}\ (\bibinfo {year} {2022})\ \Eprint {https://arxiv.org/abs/2203.14923} {arXiv:2203.14923 [hep-ex]} \BibitemShut {NoStop}%
\bibitem [{\citenamefont {Jaeckel}\ \emph {et~al.}(2016)\citenamefont {Jaeckel}, \citenamefont {Khoze},\ and\ \citenamefont {Spannowsky}}]{Jaeckel:2016jlh}%
  \BibitemOpen
  \bibfield  {author} {\bibinfo {author} {\bibfnamefont {J.}~\bibnamefont {Jaeckel}}, \bibinfo {author} {\bibfnamefont {V.~V.}\ \bibnamefont {Khoze}},\ and\ \bibinfo {author} {\bibfnamefont {M.}~\bibnamefont {Spannowsky}},\ }\href {https://doi.org/10.1103/PhysRevD.94.103519} {\bibfield  {journal} {\bibinfo  {journal} {Phys. Rev. D}\ }\textbf {\bibinfo {volume} {94}},\ \bibinfo {pages} {103519} (\bibinfo {year} {2016})},\ \Eprint {https://arxiv.org/abs/1602.03901} {arXiv:1602.03901 [hep-ph]} \BibitemShut {NoStop}%
\bibitem [{\citenamefont {Grote}\ and\ \citenamefont {Stadnik}(2019)}]{Grote:2019uvn}%
  \BibitemOpen
  \bibfield  {author} {\bibinfo {author} {\bibfnamefont {H.}~\bibnamefont {Grote}}\ and\ \bibinfo {author} {\bibfnamefont {Y.~V.}\ \bibnamefont {Stadnik}},\ }\href {https://doi.org/10.1103/PhysRevResearch.1.033187} {\bibfield  {journal} {\bibinfo  {journal} {Phys. Rev. Res.}\ }\textbf {\bibinfo {volume} {1}},\ \bibinfo {pages} {033187} (\bibinfo {year} {2019})},\ \Eprint {https://arxiv.org/abs/1906.06193} {arXiv:1906.06193 [astro-ph.IM]} \BibitemShut {NoStop}%
\bibitem [{\citenamefont {Khoze}\ and\ \citenamefont {Milne}(2022)}]{Khoze:2021uim}%
  \BibitemOpen
  \bibfield  {author} {\bibinfo {author} {\bibfnamefont {V.~V.}\ \bibnamefont {Khoze}}\ and\ \bibinfo {author} {\bibfnamefont {D.~L.}\ \bibnamefont {Milne}},\ }\href {https://doi.org/10.1016/j.physletb.2022.137044} {\bibfield  {journal} {\bibinfo  {journal} {Phys. Lett. B}\ }\textbf {\bibinfo {volume} {829}},\ \bibinfo {pages} {137044} (\bibinfo {year} {2022})},\ \Eprint {https://arxiv.org/abs/2107.02640} {arXiv:2107.02640 [hep-ph]} \BibitemShut {NoStop}%
\bibitem [{\citenamefont {Pospelov}\ \emph {et~al.}(2013)\citenamefont {Pospelov}, \citenamefont {Pustelny}, \citenamefont {Ledbetter}, \citenamefont {Jackson~Kimball}, \citenamefont {Gawlik},\ and\ \citenamefont {Budker}}]{Pospelov:2012mt}%
  \BibitemOpen
  \bibfield  {author} {\bibinfo {author} {\bibfnamefont {M.}~\bibnamefont {Pospelov}}, \bibinfo {author} {\bibfnamefont {S.}~\bibnamefont {Pustelny}}, \bibinfo {author} {\bibfnamefont {M.~P.}\ \bibnamefont {Ledbetter}}, \bibinfo {author} {\bibfnamefont {D.~F.}\ \bibnamefont {Jackson~Kimball}}, \bibinfo {author} {\bibfnamefont {W.}~\bibnamefont {Gawlik}},\ and\ \bibinfo {author} {\bibfnamefont {D.}~\bibnamefont {Budker}},\ }\href {https://doi.org/10.1103/PhysRevLett.110.021803} {\bibfield  {journal} {\bibinfo  {journal} {Phys. Rev. Lett.}\ }\textbf {\bibinfo {volume} {110}},\ \bibinfo {pages} {021803} (\bibinfo {year} {2013})},\ \Eprint {https://arxiv.org/abs/1205.6260} {arXiv:1205.6260 [hep-ph]} \BibitemShut {NoStop}%
\bibitem [{\citenamefont {Vermeulen}\ \emph {et~al.}(2021)\citenamefont {Vermeulen}, \citenamefont {Relton}, \citenamefont {Grote}, \citenamefont {Raymond}, \citenamefont {Affeldt}, \citenamefont {Bergamin}, \citenamefont {Bisht}, \citenamefont {Brinkmann}, \citenamefont {Danzmann}, \citenamefont {Doravari} \emph {et~al.}}]{vermeulen2021direct}%
  \BibitemOpen
  \bibfield  {author} {\bibinfo {author} {\bibfnamefont {S.~M.}\ \bibnamefont {Vermeulen}}, \bibinfo {author} {\bibfnamefont {P.}~\bibnamefont {Relton}}, \bibinfo {author} {\bibfnamefont {H.}~\bibnamefont {Grote}}, \bibinfo {author} {\bibfnamefont {V.}~\bibnamefont {Raymond}}, \bibinfo {author} {\bibfnamefont {C.}~\bibnamefont {Affeldt}}, \bibinfo {author} {\bibfnamefont {F.}~\bibnamefont {Bergamin}}, \bibinfo {author} {\bibfnamefont {A.}~\bibnamefont {Bisht}}, \bibinfo {author} {\bibfnamefont {M.}~\bibnamefont {Brinkmann}}, \bibinfo {author} {\bibfnamefont {K.}~\bibnamefont {Danzmann}}, \bibinfo {author} {\bibfnamefont {S.}~\bibnamefont {Doravari}}, \emph {et~al.},\ }\href@noop {} {\bibfield  {journal} {\bibinfo  {journal} {Nature}\ }\textbf {\bibinfo {volume} {600}},\ \bibinfo {pages} {424} (\bibinfo {year} {2021})}\BibitemShut {NoStop}%
\bibitem [{\citenamefont {Dai}\ \emph {et~al.}(2022)\citenamefont {Dai}, \citenamefont {Minic},\ and\ \citenamefont {Stojkovic}}]{Dai:2021boq}%
  \BibitemOpen
  \bibfield  {author} {\bibinfo {author} {\bibfnamefont {D.-C.}\ \bibnamefont {Dai}}, \bibinfo {author} {\bibfnamefont {D.}~\bibnamefont {Minic}},\ and\ \bibinfo {author} {\bibfnamefont {D.}~\bibnamefont {Stojkovic}},\ }\href {https://doi.org/10.1007/JHEP03(2022)207} {\bibfield  {journal} {\bibinfo  {journal} {JHEP}\ }\textbf {\bibinfo {volume} {03}},\ \bibinfo {pages} {207}},\ \Eprint {https://arxiv.org/abs/2105.01894} {arXiv:2105.01894 [gr-qc]} \BibitemShut {NoStop}%
\bibitem [{\citenamefont {Kibble}(1976)}]{Kibble:1976sj}%
  \BibitemOpen
  \bibfield  {author} {\bibinfo {author} {\bibfnamefont {T.~W.~B.}\ \bibnamefont {Kibble}},\ }\href {https://doi.org/10.1088/0305-4470/9/8/029} {\bibfield  {journal} {\bibinfo  {journal} {J. Phys. A}\ }\textbf {\bibinfo {volume} {9}},\ \bibinfo {pages} {1387} (\bibinfo {year} {1976})}\BibitemShut {NoStop}%
\bibitem [{\citenamefont {Vilenkin}\ and\ \citenamefont {Shellard}(2000)}]{Vilenkin:2000jqa}%
  \BibitemOpen
  \bibfield  {author} {\bibinfo {author} {\bibfnamefont {A.}~\bibnamefont {Vilenkin}}\ and\ \bibinfo {author} {\bibfnamefont {E.~P.~S.}\ \bibnamefont {Shellard}},\ }\href@noop {} {\emph {\bibinfo {title} {{Cosmic Strings and Other Topological Defects}}}}\ (\bibinfo  {publisher} {Cambridge University Press},\ \bibinfo {year} {2000})\BibitemShut {NoStop}%
\bibitem [{\citenamefont {Vachaspati}(2006)}]{Vachaspati:2006zz}%
  \BibitemOpen
  \bibfield  {author} {\bibinfo {author} {\bibfnamefont {T.}~\bibnamefont {Vachaspati}},\ }\href@noop {} {\emph {\bibinfo {title} {Kinks and Domain Walls: An Introduction to Classical and Quantum Solitons}}}\ (\bibinfo  {publisher} {Cambridge University Press},\ \bibinfo {year} {2006})\BibitemShut {NoStop}%
\bibitem [{\citenamefont {Stadnik}\ and\ \citenamefont {Flambaum}(2014)}]{Stadnik:2014cea}%
  \BibitemOpen
  \bibfield  {author} {\bibinfo {author} {\bibfnamefont {Y.~V.}\ \bibnamefont {Stadnik}}\ and\ \bibinfo {author} {\bibfnamefont {V.~V.}\ \bibnamefont {Flambaum}},\ }\href {https://doi.org/10.1103/PhysRevLett.113.151301} {\bibfield  {journal} {\bibinfo  {journal} {Phys. Rev. Lett.}\ }\textbf {\bibinfo {volume} {113}},\ \bibinfo {pages} {151301} (\bibinfo {year} {2014})},\ \Eprint {https://arxiv.org/abs/1405.5337} {arXiv:1405.5337 [hep-ph]} \BibitemShut {NoStop}%
\bibitem [{\citenamefont {Derevianko}\ and\ \citenamefont {Pospelov}(2014)}]{Derevianko:2013oaa}%
  \BibitemOpen
  \bibfield  {author} {\bibinfo {author} {\bibfnamefont {A.}~\bibnamefont {Derevianko}}\ and\ \bibinfo {author} {\bibfnamefont {M.}~\bibnamefont {Pospelov}},\ }\href {https://doi.org/10.1038/nphys3137} {\bibfield  {journal} {\bibinfo  {journal} {Nature Phys.}\ }\textbf {\bibinfo {volume} {10}},\ \bibinfo {pages} {933} (\bibinfo {year} {2014})},\ \Eprint {https://arxiv.org/abs/1311.1244} {arXiv:1311.1244 [physics.atom-ph]} \BibitemShut {NoStop}%
\bibitem [{\citenamefont {Roberts}\ \emph {et~al.}(2017)\citenamefont {Roberts}, \citenamefont {Blewitt}, \citenamefont {Dailey}, \citenamefont {Murphy}, \citenamefont {Pospelov}, \citenamefont {Rollings}, \citenamefont {Sherman}, \citenamefont {Williams},\ and\ \citenamefont {Derevianko}}]{Roberts:2017hla}%
  \BibitemOpen
  \bibfield  {author} {\bibinfo {author} {\bibfnamefont {B.~M.}\ \bibnamefont {Roberts}}, \bibinfo {author} {\bibfnamefont {G.}~\bibnamefont {Blewitt}}, \bibinfo {author} {\bibfnamefont {C.}~\bibnamefont {Dailey}}, \bibinfo {author} {\bibfnamefont {M.}~\bibnamefont {Murphy}}, \bibinfo {author} {\bibfnamefont {M.}~\bibnamefont {Pospelov}}, \bibinfo {author} {\bibfnamefont {A.}~\bibnamefont {Rollings}}, \bibinfo {author} {\bibfnamefont {J.}~\bibnamefont {Sherman}}, \bibinfo {author} {\bibfnamefont {W.}~\bibnamefont {Williams}},\ and\ \bibinfo {author} {\bibfnamefont {A.}~\bibnamefont {Derevianko}},\ }\href {https://doi.org/10.1038/s41467-017-01440-4} {\bibfield  {journal} {\bibinfo  {journal} {Nature Commun.}\ }\textbf {\bibinfo {volume} {8}},\ \bibinfo {pages} {1195} (\bibinfo {year} {2017})},\ \Eprint {https://arxiv.org/abs/1704.06844} {arXiv:1704.06844 [hep-ph]} \BibitemShut {NoStop}%
\bibitem [{\citenamefont {Heisenberg}\ \emph {et~al.}(2024)\citenamefont {Heisenberg}, \citenamefont {Maibach},\ and\ \citenamefont {Veske}}]{Heisenberg:2023urf}%
  \BibitemOpen
  \bibfield  {author} {\bibinfo {author} {\bibfnamefont {L.}~\bibnamefont {Heisenberg}}, \bibinfo {author} {\bibfnamefont {D.}~\bibnamefont {Maibach}},\ and\ \bibinfo {author} {\bibfnamefont {D.}~\bibnamefont {Veske}},\ }\href {https://doi.org/10.1103/PhysRevD.110.055037} {\bibfield  {journal} {\bibinfo  {journal} {Phys. Rev. D}\ }\textbf {\bibinfo {volume} {110}},\ \bibinfo {pages} {055037} (\bibinfo {year} {2024})},\ \Eprint {https://arxiv.org/abs/2309.05093} {arXiv:2309.05093 [gr-qc]} \BibitemShut {NoStop}%
\bibitem [{\citenamefont {Rajaraman}(1982)}]{rajaraman1982solitons}%
  \BibitemOpen
  \bibfield  {author} {\bibinfo {author} {\bibfnamefont {R.}~\bibnamefont {Rajaraman}},\ }\href@noop {} {\emph {\bibinfo {title} {Solitons and Instantons: An Introduction to Solitons and Instantons in Quantum Field Theory}}}\ (\bibinfo  {publisher} {Elsevier, New York},\ \bibinfo {year} {1982})\BibitemShut {NoStop}%
\bibitem [{\citenamefont {Manton}\ and\ \citenamefont {Sutcliffe}(2004)}]{Manton:2004tk}%
  \BibitemOpen
  \bibfield  {author} {\bibinfo {author} {\bibfnamefont {N.~S.}\ \bibnamefont {Manton}}\ and\ \bibinfo {author} {\bibfnamefont {P.}~\bibnamefont {Sutcliffe}},\ }\href {https://doi.org/10.1017/CBO9780511617034} {\emph {\bibinfo {title} {{Topological solitons}}}},\ Cambridge Monographs on Mathematical Physics\ (\bibinfo  {publisher} {Cambridge University Press},\ \bibinfo {year} {2004})\BibitemShut {NoStop}%
\bibitem [{\citenamefont {Pierce}\ \emph {et~al.}(2018)\citenamefont {Pierce}, \citenamefont {Riles},\ and\ \citenamefont {Zhao}}]{Pierce:2018xmy}%
  \BibitemOpen
  \bibfield  {author} {\bibinfo {author} {\bibfnamefont {A.}~\bibnamefont {Pierce}}, \bibinfo {author} {\bibfnamefont {K.}~\bibnamefont {Riles}},\ and\ \bibinfo {author} {\bibfnamefont {Y.}~\bibnamefont {Zhao}},\ }\href {https://doi.org/10.1103/PhysRevLett.121.061102} {\bibfield  {journal} {\bibinfo  {journal} {Phys. Rev. Lett.}\ }\textbf {\bibinfo {volume} {121}},\ \bibinfo {pages} {061102} (\bibinfo {year} {2018})},\ \Eprint {https://arxiv.org/abs/1801.10161} {arXiv:1801.10161 [hep-ph]} \BibitemShut {NoStop}%
\bibitem [{\citenamefont {Guo}\ \emph {et~al.}(2019)\citenamefont {Guo}, \citenamefont {Riles}, \citenamefont {Yang},\ and\ \citenamefont {Zhao}}]{Guo:2019ker}%
  \BibitemOpen
  \bibfield  {author} {\bibinfo {author} {\bibfnamefont {H.-K.}\ \bibnamefont {Guo}}, \bibinfo {author} {\bibfnamefont {K.}~\bibnamefont {Riles}}, \bibinfo {author} {\bibfnamefont {F.-W.}\ \bibnamefont {Yang}},\ and\ \bibinfo {author} {\bibfnamefont {Y.}~\bibnamefont {Zhao}},\ }\href {https://doi.org/10.1038/s42005-019-0255-0} {\bibfield  {journal} {\bibinfo  {journal} {Commun. Phys.}\ }\textbf {\bibinfo {volume} {2}},\ \bibinfo {pages} {155} (\bibinfo {year} {2019})},\ \Eprint {https://arxiv.org/abs/1905.04316} {arXiv:1905.04316 [hep-ph]} \BibitemShut {NoStop}%
\bibitem [{\citenamefont {Abbott}\ \emph {et~al.}(2022)\citenamefont {Abbott} \emph {et~al.}}]{LIGOScientific:2021ffg}%
  \BibitemOpen
  \bibfield  {author} {\bibinfo {author} {\bibfnamefont {R.}~\bibnamefont {Abbott}} \emph {et~al.} (\bibinfo {collaboration} {LIGO Scientific, KAGRA, Virgo}),\ }\href {https://doi.org/10.1103/PhysRevD.105.063030} {\bibfield  {journal} {\bibinfo  {journal} {Phys. Rev. D}\ }\textbf {\bibinfo {volume} {105}},\ \bibinfo {pages} {063030} (\bibinfo {year} {2022})},\ \bibinfo {note} {[Erratum: Phys.Rev.D 109, 089902 (2024)]},\ \Eprint {https://arxiv.org/abs/2105.13085} {arXiv:2105.13085 [astro-ph.CO]} \BibitemShut {NoStop}%
\bibitem [{\citenamefont {Abac}\ \emph {et~al.}(2026)\citenamefont {Abac} \emph {et~al.}}]{theligoscientificcollaboration2025directmultimodeldarkmattersearch}%
  \BibitemOpen
  \bibfield  {author} {\bibinfo {author} {\bibfnamefont {A.~G.}\ \bibnamefont {Abac}} \emph {et~al.} (\bibinfo {collaboration} {LIGO Scientific, VIRGO, KAGRA}),\ }\href {https://doi.org/10.1103/qdnd-rsnh} {\bibinfo {title} {{Direct Multimodel Dark Matter Search with Gravitational-Wave Interferometers Using Data from the First Part of the Fourth LIGO-Virgo-KAGRA Observing Run}}} (\bibinfo {year} {2026}),\ \Eprint {https://arxiv.org/abs/2510.27022} {arXiv:2510.27022 [astro-ph.CO]} \BibitemShut {NoStop}%
\bibitem [{\citenamefont {Fell}\ \emph {et~al.}(2023)\citenamefont {Fell}, \citenamefont {Heisenberg},\ and\ \citenamefont {Veske}}]{Fell:2023mtf}%
  \BibitemOpen
  \bibfield  {author} {\bibinfo {author} {\bibfnamefont {S.~D.~B.}\ \bibnamefont {Fell}}, \bibinfo {author} {\bibfnamefont {L.}~\bibnamefont {Heisenberg}},\ and\ \bibinfo {author} {\bibfnamefont {D.}~\bibnamefont {Veske}},\ }\href {https://doi.org/10.1103/PhysRevD.108.083010} {\bibfield  {journal} {\bibinfo  {journal} {Phys. Rev. D}\ }\textbf {\bibinfo {volume} {108}},\ \bibinfo {pages} {083010} (\bibinfo {year} {2023})},\ \Eprint {https://arxiv.org/abs/2304.14129} {arXiv:2304.14129 [gr-qc]} \BibitemShut {NoStop}%
\bibitem [{\citenamefont {Bhattacharya}\ \emph {et~al.}(2023)\citenamefont {Bhattacharya}, \citenamefont {Dasgupta}, \citenamefont {Laha},\ and\ \citenamefont {Ray}}]{Bhattacharya:2023stq}%
  \BibitemOpen
  \bibfield  {author} {\bibinfo {author} {\bibfnamefont {S.}~\bibnamefont {Bhattacharya}}, \bibinfo {author} {\bibfnamefont {B.}~\bibnamefont {Dasgupta}}, \bibinfo {author} {\bibfnamefont {R.}~\bibnamefont {Laha}},\ and\ \bibinfo {author} {\bibfnamefont {A.}~\bibnamefont {Ray}},\ }\href {https://doi.org/10.1103/PhysRevLett.131.091401} {\bibfield  {journal} {\bibinfo  {journal} {Phys. Rev. Lett.}\ }\textbf {\bibinfo {volume} {131}},\ \bibinfo {pages} {091401} (\bibinfo {year} {2023})},\ \Eprint {https://arxiv.org/abs/2302.07898} {arXiv:2302.07898 [hep-ph]} \BibitemShut {NoStop}%
\bibitem [{\citenamefont {Dasgupta}\ \emph {et~al.}(2021)\citenamefont {Dasgupta}, \citenamefont {Laha},\ and\ \citenamefont {Ray}}]{Dasgupta:2020mqg}%
  \BibitemOpen
  \bibfield  {author} {\bibinfo {author} {\bibfnamefont {B.}~\bibnamefont {Dasgupta}}, \bibinfo {author} {\bibfnamefont {R.}~\bibnamefont {Laha}},\ and\ \bibinfo {author} {\bibfnamefont {A.}~\bibnamefont {Ray}},\ }\href {https://doi.org/10.1103/PhysRevLett.126.141105} {\bibfield  {journal} {\bibinfo  {journal} {Phys. Rev. Lett.}\ }\textbf {\bibinfo {volume} {126}},\ \bibinfo {pages} {141105} (\bibinfo {year} {2021})},\ \Eprint {https://arxiv.org/abs/2009.01825} {arXiv:2009.01825 [astro-ph.HE]} \BibitemShut {NoStop}%
\bibitem [{\citenamefont {Baryakhtar}\ \emph {et~al.}(2017)\citenamefont {Baryakhtar}, \citenamefont {Lasenby},\ and\ \citenamefont {Teo}}]{Baryakhtar:2017ngi}%
  \BibitemOpen
  \bibfield  {author} {\bibinfo {author} {\bibfnamefont {M.}~\bibnamefont {Baryakhtar}}, \bibinfo {author} {\bibfnamefont {R.}~\bibnamefont {Lasenby}},\ and\ \bibinfo {author} {\bibfnamefont {M.}~\bibnamefont {Teo}},\ }\href {https://doi.org/10.1103/PhysRevD.96.035019} {\bibfield  {journal} {\bibinfo  {journal} {Phys. Rev. D}\ }\textbf {\bibinfo {volume} {96}},\ \bibinfo {pages} {035019} (\bibinfo {year} {2017})},\ \Eprint {https://arxiv.org/abs/1704.05081} {arXiv:1704.05081 [hep-ph]} \BibitemShut {NoStop}%
\bibitem [{\citenamefont {Maselli}\ \emph {et~al.}(2020)\citenamefont {Maselli}, \citenamefont {Franchini}, \citenamefont {Gualtieri},\ and\ \citenamefont {Sotiriou}}]{Maselli:2020zgv}%
  \BibitemOpen
  \bibfield  {author} {\bibinfo {author} {\bibfnamefont {A.}~\bibnamefont {Maselli}}, \bibinfo {author} {\bibfnamefont {N.}~\bibnamefont {Franchini}}, \bibinfo {author} {\bibfnamefont {L.}~\bibnamefont {Gualtieri}},\ and\ \bibinfo {author} {\bibfnamefont {T.~P.}\ \bibnamefont {Sotiriou}},\ }\href {https://doi.org/10.1103/PhysRevLett.125.141101} {\bibfield  {journal} {\bibinfo  {journal} {Phys. Rev. Lett.}\ }\textbf {\bibinfo {volume} {125}},\ \bibinfo {pages} {141101} (\bibinfo {year} {2020})},\ \Eprint {https://arxiv.org/abs/2004.11895} {arXiv:2004.11895 [gr-qc]} \BibitemShut {NoStop}%
\bibitem [{\citenamefont {Maselli}\ \emph {et~al.}(2022)\citenamefont {Maselli}, \citenamefont {Franchini}, \citenamefont {Gualtieri}, \citenamefont {Sotiriou}, \citenamefont {Barsanti},\ and\ \citenamefont {Pani}}]{Maselli:2021men}%
  \BibitemOpen
  \bibfield  {author} {\bibinfo {author} {\bibfnamefont {A.}~\bibnamefont {Maselli}}, \bibinfo {author} {\bibfnamefont {N.}~\bibnamefont {Franchini}}, \bibinfo {author} {\bibfnamefont {L.}~\bibnamefont {Gualtieri}}, \bibinfo {author} {\bibfnamefont {T.~P.}\ \bibnamefont {Sotiriou}}, \bibinfo {author} {\bibfnamefont {S.}~\bibnamefont {Barsanti}},\ and\ \bibinfo {author} {\bibfnamefont {P.}~\bibnamefont {Pani}},\ }\href {https://doi.org/10.1038/s41550-021-01589-5} {\bibfield  {journal} {\bibinfo  {journal} {Nature Astron.}\ }\textbf {\bibinfo {volume} {6}},\ \bibinfo {pages} {464} (\bibinfo {year} {2022})},\ \Eprint {https://arxiv.org/abs/2106.11325} {arXiv:2106.11325 [gr-qc]} \BibitemShut {NoStop}%
\bibitem [{\citenamefont {Bartolo}\ \emph {et~al.}(2019)\citenamefont {Bartolo}, \citenamefont {De~Luca}, \citenamefont {Franciolini}, \citenamefont {Lewis}, \citenamefont {Peloso},\ and\ \citenamefont {Riotto}}]{Bartolo:2018evs}%
  \BibitemOpen
  \bibfield  {author} {\bibinfo {author} {\bibfnamefont {N.}~\bibnamefont {Bartolo}}, \bibinfo {author} {\bibfnamefont {V.}~\bibnamefont {De~Luca}}, \bibinfo {author} {\bibfnamefont {G.}~\bibnamefont {Franciolini}}, \bibinfo {author} {\bibfnamefont {A.}~\bibnamefont {Lewis}}, \bibinfo {author} {\bibfnamefont {M.}~\bibnamefont {Peloso}},\ and\ \bibinfo {author} {\bibfnamefont {A.}~\bibnamefont {Riotto}},\ }\href {https://doi.org/10.1103/PhysRevLett.122.211301} {\bibfield  {journal} {\bibinfo  {journal} {Phys. Rev. Lett.}\ }\textbf {\bibinfo {volume} {122}},\ \bibinfo {pages} {211301} (\bibinfo {year} {2019})},\ \Eprint {https://arxiv.org/abs/1810.12218} {arXiv:1810.12218 [astro-ph.CO]} \BibitemShut {NoStop}%
\bibitem [{\citenamefont {Abac}\ \emph {et~al.}(2025{\natexlab{a}})\citenamefont {Abac} \emph {et~al.}}]{LIGOScientific:2025rsn}%
  \BibitemOpen
  \bibfield  {author} {\bibinfo {author} {\bibfnamefont {A.~G.}\ \bibnamefont {Abac}} \emph {et~al.} (\bibinfo {collaboration} {LIGO Scientific, VIRGO, KAGRA}),\ }\href {https://doi.org/10.3847/2041-8213/ae0c9c} {\bibfield  {journal} {\bibinfo  {journal} {Astrophys. J. Lett.}\ }\textbf {\bibinfo {volume} {993}},\ \bibinfo {pages} {L25} (\bibinfo {year} {2025}{\natexlab{a}})},\ \Eprint {https://arxiv.org/abs/2507.08219} {arXiv:2507.08219 [astro-ph.HE]} \BibitemShut {NoStop}%
\bibitem [{\citenamefont {Aasi}\ \emph {et~al.}(2015)\citenamefont {Aasi} \emph {et~al.}}]{LIGOScientific:2014pky}%
  \BibitemOpen
  \bibfield  {author} {\bibinfo {author} {\bibfnamefont {J.}~\bibnamefont {Aasi}} \emph {et~al.} (\bibinfo {collaboration} {LIGO Scientific}),\ }\href {https://doi.org/10.1088/0264-9381/32/7/074001} {\bibfield  {journal} {\bibinfo  {journal} {Class. Quant. Grav.}\ }\textbf {\bibinfo {volume} {32}},\ \bibinfo {pages} {074001} (\bibinfo {year} {2015})},\ \Eprint {https://arxiv.org/abs/1411.4547} {arXiv:1411.4547 [gr-qc]} \BibitemShut {NoStop}%
\bibitem [{\citenamefont {Li}\ and\ \citenamefont {Fan}(2025)}]{Li:2025pyo}%
  \BibitemOpen
  \bibfield  {author} {\bibinfo {author} {\bibfnamefont {G.-P.}\ \bibnamefont {Li}}\ and\ \bibinfo {author} {\bibfnamefont {X.-L.}\ \bibnamefont {Fan}},\ }\href@noop {} {\bibinfo {title} {{The Hierarchical Merger Scenario for GW231123}}} (\bibinfo {year} {2025}),\ \Eprint {https://arxiv.org/abs/2509.08298} {arXiv:2509.08298 [astro-ph.HE]} \BibitemShut {NoStop}%
\bibitem [{\citenamefont {Passenger}\ \emph {et~al.}(2025)\citenamefont {Passenger}, \citenamefont {Banagiri}, \citenamefont {Thrane}, \citenamefont {Lasky}, \citenamefont {Borchers}, \citenamefont {Fishbach},\ and\ \citenamefont {Ye}}]{Passenger:2025acb}%
  \BibitemOpen
  \bibfield  {author} {\bibinfo {author} {\bibfnamefont {L.}~\bibnamefont {Passenger}}, \bibinfo {author} {\bibfnamefont {S.}~\bibnamefont {Banagiri}}, \bibinfo {author} {\bibfnamefont {E.}~\bibnamefont {Thrane}}, \bibinfo {author} {\bibfnamefont {P.~D.}\ \bibnamefont {Lasky}}, \bibinfo {author} {\bibfnamefont {A.}~\bibnamefont {Borchers}}, \bibinfo {author} {\bibfnamefont {M.}~\bibnamefont {Fishbach}},\ and\ \bibinfo {author} {\bibfnamefont {C.~S.}\ \bibnamefont {Ye}},\ }\href@noop {} {\bibinfo {title} {{Is GW231123 a hierarchical merger?}}} (\bibinfo {year} {2025}),\ \Eprint {https://arxiv.org/abs/2510.14363} {arXiv:2510.14363 [astro-ph.HE]} \BibitemShut {NoStop}%
\bibitem [{\citenamefont {Paiella}\ \emph {et~al.}(2025)\citenamefont {Paiella}, \citenamefont {Ugolini}, \citenamefont {Spera}, \citenamefont {Branchesi},\ and\ \citenamefont {Sedda}}]{Paiella:2025qld}%
  \BibitemOpen
  \bibfield  {author} {\bibinfo {author} {\bibfnamefont {L.}~\bibnamefont {Paiella}}, \bibinfo {author} {\bibfnamefont {C.}~\bibnamefont {Ugolini}}, \bibinfo {author} {\bibfnamefont {M.}~\bibnamefont {Spera}}, \bibinfo {author} {\bibfnamefont {M.}~\bibnamefont {Branchesi}},\ and\ \bibinfo {author} {\bibfnamefont {M.~A.}\ \bibnamefont {Sedda}},\ }\href {https://doi.org/10.3847/2041-8213/ae1447} {\bibfield  {journal} {\bibinfo  {journal} {Astrophys. J. Lett.}\ }\textbf {\bibinfo {volume} {994}},\ \bibinfo {pages} {L54} (\bibinfo {year} {2025})},\ \Eprint {https://arxiv.org/abs/2509.10609} {arXiv:2509.10609 [astro-ph.GA]} \BibitemShut {NoStop}%
\bibitem [{\citenamefont {Cuceu}\ \emph {et~al.}(2026)\citenamefont {Cuceu}, \citenamefont {Bizouard}, \citenamefont {Christensen},\ and\ \citenamefont {Sakellariadou}}]{Cuceu:2025fzi}%
  \BibitemOpen
  \bibfield  {author} {\bibinfo {author} {\bibfnamefont {I.}~\bibnamefont {Cuceu}}, \bibinfo {author} {\bibfnamefont {M.~A.}\ \bibnamefont {Bizouard}}, \bibinfo {author} {\bibfnamefont {N.}~\bibnamefont {Christensen}},\ and\ \bibinfo {author} {\bibfnamefont {M.}~\bibnamefont {Sakellariadou}},\ }\href {https://doi.org/10.1103/zd8m-tzxd} {\bibfield  {journal} {\bibinfo  {journal} {Phys. Rev. D}\ }\textbf {\bibinfo {volume} {113}},\ \bibinfo {pages} {L021302} (\bibinfo {year} {2026})},\ \Eprint {https://arxiv.org/abs/2507.20778} {arXiv:2507.20778 [gr-qc]} \BibitemShut {NoStop}%
\bibitem [{\citenamefont {Abbott}\ \emph {et~al.}(2020)\citenamefont {Abbott} \emph {et~al.}}]{LIGOScientific:2020iuh}%
  \BibitemOpen
  \bibfield  {author} {\bibinfo {author} {\bibfnamefont {R.}~\bibnamefont {Abbott}} \emph {et~al.} (\bibinfo {collaboration} {LIGO Scientific, Virgo}),\ }\href {https://doi.org/10.1103/PhysRevLett.125.101102} {\bibfield  {journal} {\bibinfo  {journal} {Phys. Rev. Lett.}\ }\textbf {\bibinfo {volume} {125}},\ \bibinfo {pages} {101102} (\bibinfo {year} {2020})},\ \Eprint {https://arxiv.org/abs/2009.01075} {arXiv:2009.01075 [gr-qc]} \BibitemShut {NoStop}%
\bibitem [{\citenamefont {Uzan}(2011)}]{Uzan:2010pm}%
  \BibitemOpen
  \bibfield  {author} {\bibinfo {author} {\bibfnamefont {J.-P.}\ \bibnamefont {Uzan}},\ }\href {https://doi.org/10.12942/lrr-2011-2} {\bibfield  {journal} {\bibinfo  {journal} {Living Rev. Rel.}\ }\textbf {\bibinfo {volume} {14}},\ \bibinfo {pages} {2} (\bibinfo {year} {2011})},\ \Eprint {https://arxiv.org/abs/1009.5514} {arXiv:1009.5514 [astro-ph.CO]} \BibitemShut {NoStop}%
\bibitem [{sup()}]{supp}%
  \BibitemOpen
  \href@noop {} {}\bibinfo {note} {See Supplemental Material at [URL will be inserted by publisher] for details on detector response channels, data conditioning, prior choices, BBH analysis, robustness checks, and posterior corner plots, which includes Refs.~\cite{Varma:2019csw,LIGOScientific:2021usb,KAGRA:2021vkt}}\BibitemShut {NoStop}%
\bibitem [{\citenamefont {Varma}\ \emph {et~al.}(2019)\citenamefont {Varma}, \citenamefont {Field}, \citenamefont {Scheel}, \citenamefont {Blackman}, \citenamefont {Gerosa}, \citenamefont {Stein}, \citenamefont {Kidder},\ and\ \citenamefont {Pfeiffer}}]{Varma:2019csw}%
  \BibitemOpen
  \bibfield  {author} {\bibinfo {author} {\bibfnamefont {V.}~\bibnamefont {Varma}}, \bibinfo {author} {\bibfnamefont {S.~E.}\ \bibnamefont {Field}}, \bibinfo {author} {\bibfnamefont {M.~A.}\ \bibnamefont {Scheel}}, \bibinfo {author} {\bibfnamefont {J.}~\bibnamefont {Blackman}}, \bibinfo {author} {\bibfnamefont {D.}~\bibnamefont {Gerosa}}, \bibinfo {author} {\bibfnamefont {L.~C.}\ \bibnamefont {Stein}}, \bibinfo {author} {\bibfnamefont {L.~E.}\ \bibnamefont {Kidder}},\ and\ \bibinfo {author} {\bibfnamefont {H.~P.}\ \bibnamefont {Pfeiffer}},\ }\href {https://doi.org/10.1103/PhysRevResearch.1.033015} {\bibfield  {journal} {\bibinfo  {journal} {Phys. Rev. Research.}\ }\textbf {\bibinfo {volume} {1}},\ \bibinfo {pages} {033015} (\bibinfo {year} {2019})},\ \Eprint {https://arxiv.org/abs/1905.09300} {arXiv:1905.09300 [gr-qc]} \BibitemShut {NoStop}%
\bibitem [{\citenamefont {Abbott}\ \emph {et~al.}(2024)\citenamefont {Abbott} \emph {et~al.}}]{LIGOScientific:2021usb}%
  \BibitemOpen
  \bibfield  {author} {\bibinfo {author} {\bibfnamefont {R.}~\bibnamefont {Abbott}} \emph {et~al.} (\bibinfo {collaboration} {LIGO Scientific, VIRGO}),\ }\href {https://doi.org/10.1103/PhysRevD.109.022001} {\bibfield  {journal} {\bibinfo  {journal} {Phys. Rev. D}\ }\textbf {\bibinfo {volume} {109}},\ \bibinfo {pages} {022001} (\bibinfo {year} {2024})},\ \Eprint {https://arxiv.org/abs/2108.01045} {arXiv:2108.01045 [gr-qc]} \BibitemShut {NoStop}%
\bibitem [{\citenamefont {Abbott}\ \emph {et~al.}(2023)\citenamefont {Abbott} \emph {et~al.}}]{KAGRA:2021vkt}%
  \BibitemOpen
  \bibfield  {author} {\bibinfo {author} {\bibfnamefont {R.}~\bibnamefont {Abbott}} \emph {et~al.} (\bibinfo {collaboration} {KAGRA, VIRGO, LIGO Scientific}),\ }\href {https://doi.org/10.1103/PhysRevX.13.041039} {\bibfield  {journal} {\bibinfo  {journal} {Phys. Rev. X}\ }\textbf {\bibinfo {volume} {13}},\ \bibinfo {pages} {041039} (\bibinfo {year} {2023})},\ \Eprint {https://arxiv.org/abs/2111.03606} {arXiv:2111.03606 [gr-qc]} \BibitemShut {NoStop}%
\bibitem [{\citenamefont {Abac}\ \emph {et~al.}(2025{\natexlab{b}})\citenamefont {Abac} \emph {et~al.}}]{LIGOScientific:2025snk}%
  \BibitemOpen
  \bibfield  {author} {\bibinfo {author} {\bibfnamefont {A.~G.}\ \bibnamefont {Abac}} \emph {et~al.} (\bibinfo {collaboration} {LIGO Scientific, VIRGO, KAGRA}),\ }\href@noop {} {\bibinfo {title} {{Open Data from LIGO, Virgo, and KAGRA through the First Part of the Fourth Observing Run}}} (\bibinfo {year} {2025}{\natexlab{b}}),\ \Eprint {https://arxiv.org/abs/2508.18079} {arXiv:2508.18079 [gr-qc]} \BibitemShut {NoStop}%
\bibitem [{\citenamefont {Ashton}\ \emph {et~al.}(2019)\citenamefont {Ashton} \emph {et~al.}}]{Ashton:2018jfp}%
  \BibitemOpen
  \bibfield  {author} {\bibinfo {author} {\bibfnamefont {G.}~\bibnamefont {Ashton}} \emph {et~al.},\ }\href {https://doi.org/10.3847/1538-4365/ab06fc} {\bibfield  {journal} {\bibinfo  {journal} {Astrophys. J. Suppl.}\ }\textbf {\bibinfo {volume} {241}},\ \bibinfo {pages} {27} (\bibinfo {year} {2019})},\ \Eprint {https://arxiv.org/abs/1811.02042} {arXiv:1811.02042 [astro-ph.IM]} \BibitemShut {NoStop}%
\bibitem [{\citenamefont {Speagle}(2020)}]{Speagle2020Dynesty}%
  \BibitemOpen
  \bibfield  {author} {\bibinfo {author} {\bibfnamefont {J.~S.}\ \bibnamefont {Speagle}},\ }\href {https://doi.org/10.1093/mnras/staa278} {\bibfield  {journal} {\bibinfo  {journal} {Monthly Notices of the Royal Astronomical Society}\ }\textbf {\bibinfo {volume} {493}},\ \bibinfo {pages} {3132} (\bibinfo {year} {2020})},\ \Eprint {https://arxiv.org/abs/1904.02180} {arXiv:1904.02180 [astro-ph.IM]} \BibitemShut {NoStop}%
\bibitem [{\citenamefont {Dankovsky}\ \emph {et~al.}(2024)\citenamefont {Dankovsky}, \citenamefont {Babichev}, \citenamefont {Gorbunov}, \citenamefont {Ramazanov},\ and\ \citenamefont {Vikman}}]{Dankovsky:2024zvs}%
  \BibitemOpen
  \bibfield  {author} {\bibinfo {author} {\bibfnamefont {I.}~\bibnamefont {Dankovsky}}, \bibinfo {author} {\bibfnamefont {E.}~\bibnamefont {Babichev}}, \bibinfo {author} {\bibfnamefont {D.}~\bibnamefont {Gorbunov}}, \bibinfo {author} {\bibfnamefont {S.}~\bibnamefont {Ramazanov}},\ and\ \bibinfo {author} {\bibfnamefont {A.}~\bibnamefont {Vikman}},\ }\href {https://doi.org/10.1088/1475-7516/2024/09/047} {\bibfield  {journal} {\bibinfo  {journal} {JCAP}\ }\textbf {\bibinfo {volume} {09}},\ \bibinfo {pages} {047}},\ \Eprint {https://arxiv.org/abs/2406.17053} {arXiv:2406.17053 [astro-ph.CO]} \BibitemShut {NoStop}%
\bibitem [{\citenamefont {Øyvind Christiansen}\ \emph {et~al.}(2026)\citenamefont {Øyvind Christiansen}, \citenamefont {Adamek},\ and\ \citenamefont {Kunz}}]{christiansen2026cosmicstringsdomainwalls}%
  \BibitemOpen
  \bibfield  {author} {\bibinfo {author} {\bibnamefont {Øyvind Christiansen}}, \bibinfo {author} {\bibfnamefont {J.}~\bibnamefont {Adamek}},\ and\ \bibinfo {author} {\bibfnamefont {M.}~\bibnamefont {Kunz}},\ }\href {https://arxiv.org/abs/2601.15234} {\bibinfo {title} {Cosmic strings, domain walls and environment-dependent clustering}} (\bibinfo {year} {2026}),\ \Eprint {https://arxiv.org/abs/2601.15234} {arXiv:2601.15234 [astro-ph.CO]} \BibitemShut {NoStop}%
\bibitem [{\citenamefont {Hinterbichler}\ and\ \citenamefont {Khoury}(2010)}]{Hinterbichler_2010}%
  \BibitemOpen
  \bibfield  {author} {\bibinfo {author} {\bibfnamefont {K.}~\bibnamefont {Hinterbichler}}\ and\ \bibinfo {author} {\bibfnamefont {J.}~\bibnamefont {Khoury}},\ }\bibfield  {journal} {\bibinfo  {journal} {Physical Review Letters}\ }\textbf {\bibinfo {volume} {104}},\ \href {https://doi.org/10.1103/physrevlett.104.231301} {10.1103/physrevlett.104.231301} (\bibinfo {year} {2010})\BibitemShut {NoStop}%
\bibitem [{\citenamefont {Hinterbichler}\ \emph {et~al.}(2011)\citenamefont {Hinterbichler}, \citenamefont {Khoury}, \citenamefont {Levy},\ and\ \citenamefont {Matas}}]{Hinterbichler_2011}%
  \BibitemOpen
  \bibfield  {author} {\bibinfo {author} {\bibfnamefont {K.}~\bibnamefont {Hinterbichler}}, \bibinfo {author} {\bibfnamefont {J.}~\bibnamefont {Khoury}}, \bibinfo {author} {\bibfnamefont {A.}~\bibnamefont {Levy}},\ and\ \bibinfo {author} {\bibfnamefont {A.}~\bibnamefont {Matas}},\ }\bibfield  {journal} {\bibinfo  {journal} {Physical Review D}\ }\textbf {\bibinfo {volume} {84}},\ \href {https://doi.org/10.1103/physrevd.84.103521} {10.1103/physrevd.84.103521} (\bibinfo {year} {2011})\BibitemShut {NoStop}%
\bibitem [{\citenamefont {Roberts}\ \emph {et~al.}(2020)\citenamefont {Roberts} \emph {et~al.}}]{Roberts:2019sfo}%
  \BibitemOpen
  \bibfield  {author} {\bibinfo {author} {\bibfnamefont {B.~M.}\ \bibnamefont {Roberts}} \emph {et~al.},\ }\href {https://doi.org/10.1088/1367-2630/abaace} {\bibfield  {journal} {\bibinfo  {journal} {New J. Phys.}\ }\textbf {\bibinfo {volume} {22}},\ \bibinfo {pages} {093010} (\bibinfo {year} {2020})},\ \Eprint {https://arxiv.org/abs/1907.02661} {arXiv:1907.02661 [astro-ph.CO]} \BibitemShut {NoStop}%
\bibitem [{\citenamefont {Avelino}\ and\ \citenamefont {Sousa}(2015)}]{Avelino:2014xda}%
  \BibitemOpen
  \bibfield  {author} {\bibinfo {author} {\bibfnamefont {P.~P.}\ \bibnamefont {Avelino}}\ and\ \bibinfo {author} {\bibfnamefont {L.}~\bibnamefont {Sousa}},\ }\href {https://doi.org/10.3390/universe1010006} {\bibfield  {journal} {\bibinfo  {journal} {Universe}\ }\textbf {\bibinfo {volume} {1}},\ \bibinfo {pages} {6} (\bibinfo {year} {2015})},\ \Eprint {https://arxiv.org/abs/1404.3419} {arXiv:1404.3419 [astro-ph.CO]} \BibitemShut {NoStop}%
\bibitem [{\citenamefont {Zeldovich}\ \emph {et~al.}(1974)\citenamefont {Zeldovich}, \citenamefont {Kobzarev},\ and\ \citenamefont {Okun}}]{Zeldovich:1974uw}%
  \BibitemOpen
  \bibfield  {author} {\bibinfo {author} {\bibfnamefont {Y.~B.}\ \bibnamefont {Zeldovich}}, \bibinfo {author} {\bibfnamefont {I.~Y.}\ \bibnamefont {Kobzarev}},\ and\ \bibinfo {author} {\bibfnamefont {L.~B.}\ \bibnamefont {Okun}},\ }\href@noop {} {\bibfield  {journal} {\bibinfo  {journal} {Zh. Eksp. Teor. Fiz.}\ }\textbf {\bibinfo {volume} {67}},\ \bibinfo {pages} {3} (\bibinfo {year} {1974})}\BibitemShut {NoStop}%
\bibitem [{\citenamefont {Saikawa}(2017)}]{Saikawa_2017}%
  \BibitemOpen
  \bibfield  {author} {\bibinfo {author} {\bibfnamefont {K.}~\bibnamefont {Saikawa}},\ }\href {https://doi.org/10.3390/universe3020040} {\bibfield  {journal} {\bibinfo  {journal} {Universe}\ }\textbf {\bibinfo {volume} {3}},\ \bibinfo {pages} {40} (\bibinfo {year} {2017})}\BibitemShut {NoStop}%
\bibitem [{\citenamefont {Burrage}\ \emph {et~al.}(2016)\citenamefont {Burrage}, \citenamefont {Kuribayashi-Coleman}, \citenamefont {Stevenson},\ and\ \citenamefont {Thrussell}}]{Burrage_2016}%
  \BibitemOpen
  \bibfield  {author} {\bibinfo {author} {\bibfnamefont {C.}~\bibnamefont {Burrage}}, \bibinfo {author} {\bibfnamefont {A.}~\bibnamefont {Kuribayashi-Coleman}}, \bibinfo {author} {\bibfnamefont {J.}~\bibnamefont {Stevenson}},\ and\ \bibinfo {author} {\bibfnamefont {B.}~\bibnamefont {Thrussell}},\ }\href {https://doi.org/10.1088/1475-7516/2016/12/041} {\bibfield  {journal} {\bibinfo  {journal} {Journal of Cosmology and Astroparticle Physics}\ }\textbf {\bibinfo {volume} {2016}}\bibinfo  {number} { (12)},\ \bibinfo {pages} {041–041}}\BibitemShut {NoStop}%
\bibitem [{\citenamefont {Panda}\ \emph {et~al.}(2024)\citenamefont {Panda}, \citenamefont {Tao}, \citenamefont {Ceja}, \citenamefont {Khoury}, \citenamefont {Tino},\ and\ \citenamefont {M{\"u}ller}}]{Panda:2023nir}%
  \BibitemOpen
\bibfield  {number} {  }\bibfield  {author} {\bibinfo {author} {\bibfnamefont {C.~D.}\ \bibnamefont {Panda}}, \bibinfo {author} {\bibfnamefont {M.~J.}\ \bibnamefont {Tao}}, \bibinfo {author} {\bibfnamefont {M.}~\bibnamefont {Ceja}}, \bibinfo {author} {\bibfnamefont {J.}~\bibnamefont {Khoury}}, \bibinfo {author} {\bibfnamefont {G.~M.}\ \bibnamefont {Tino}},\ and\ \bibinfo {author} {\bibfnamefont {H.}~\bibnamefont {M{\"u}ller}},\ }\href {https://doi.org/10.1038/s41586-024-07561-3} {\bibfield  {journal} {\bibinfo  {journal} {Nature}\ }\textbf {\bibinfo {volume} {631}},\ \bibinfo {pages} {515} (\bibinfo {year} {2024})},\ \Eprint {https://arxiv.org/abs/2310.01344} {arXiv:2310.01344 [physics.atom-ph]} \BibitemShut {NoStop}%
\bibitem [{\citenamefont {Cronenberg}\ \emph {et~al.}(2018)\citenamefont {Cronenberg}, \citenamefont {Brax}, \citenamefont {Filter}, \citenamefont {Geltenbort}, \citenamefont {Jenke}, \citenamefont {Pignol}, \citenamefont {Pitschmann}, \citenamefont {Thalhammer},\ and\ \citenamefont {Abele}}]{Cronenberg_2018}%
  \BibitemOpen
  \bibfield  {author} {\bibinfo {author} {\bibfnamefont {G.}~\bibnamefont {Cronenberg}}, \bibinfo {author} {\bibfnamefont {P.}~\bibnamefont {Brax}}, \bibinfo {author} {\bibfnamefont {H.}~\bibnamefont {Filter}}, \bibinfo {author} {\bibfnamefont {P.}~\bibnamefont {Geltenbort}}, \bibinfo {author} {\bibfnamefont {T.}~\bibnamefont {Jenke}}, \bibinfo {author} {\bibfnamefont {G.}~\bibnamefont {Pignol}}, \bibinfo {author} {\bibfnamefont {M.}~\bibnamefont {Pitschmann}}, \bibinfo {author} {\bibfnamefont {M.}~\bibnamefont {Thalhammer}},\ and\ \bibinfo {author} {\bibfnamefont {H.}~\bibnamefont {Abele}},\ }\href {https://doi.org/10.1038/s41567-018-0205-x} {\bibfield  {journal} {\bibinfo  {journal} {Nature Physics}\ }\textbf {\bibinfo {volume} {14}},\ \bibinfo {pages} {1022–1026} (\bibinfo {year} {2018})}\BibitemShut {NoStop}%
\end{thebibliography}
\end{document}


\title{Supplemental Material for ``Short Gravitational-Wave Transients as Probes of Cosmic Domain Walls''}
\author{T\"{o}re Boybeyi}
\affiliation{School of Physics \& Astronomy, University of Minnesota, Minneapolis, 55455, MN, USA}
\author{Do\u{g}a Veske}
\affiliation{Fizik B\"ol\"um\"u, Orta Do\u{g}u Teknik \"Universitesi, \c{C}ankaya, Ankara 06800, T\"urkiye}
\affiliation{Uzay ve H{\i}zland{\i}r{\i}c{\i} Teknolojileri Uygulama ve Ara\c{s}t{\i}rma Merkezi, Orta Do\u{g}u Teknik \"Universitesi, \c{C}ankaya, Ankara 06800, T\"urkiye}
\affiliation{Columbia Astrophysics Laboratory, Columbia University in the City of New York, New York, NY 10027, USA}
\author{David Maibach}
\affiliation{Institute for Theoretical Physics, University of Heidelberg, Philosophenweg 16, D-69120 Heidelberg, Germany}
\maketitle

\section{Detector response channels}

Let $\hat{\bm x}$ and $\hat{\bm y}$ denote the unit vectors along the two arms and let $L$ be the arm length.
We evaluate the wall coordinate at the optical elements as
$u_{\rm BS}\equiv u(\bm 0,t)$, $u_X\equiv u(L\hat{\bm x},t)$, and $u_Y\equiv u(L\hat{\bm y},t)$.

The total phase shift from dimensional effects is given by~\cite{Heisenberg:2023urf}
\begin{align}
    \Delta\varphi_{\rm size}(t) \simeq \frac{2\pi}{\lambda} g_{\rm DW} \Big[ &A_{\rm BS} \mathcal{S}(u_{\rm BS}) \nonumber \\
    &+ A_{\rm M} \left( \mathcal{S}(u_{Y}) - \mathcal{S}(u_{X}) \right) \Big].
\label{eq:phase_dimensional}
\end{align}
For Advanced LIGO, the laser wavelength is $\lambda = 1064$~nm~\cite{LIGOScientific:2014pky}. The coefficients $A_{\rm BS} \approx \sqrt{2} n l_{\rm BS}/N_{\rm eff}$ and $A_{\rm M} \approx w_{\rm M}$ are determined using the refractive index of fused silica $n \approx 1.45$, beam splitter thickness $l_{\rm BS} \approx 0.06$~m, mirror thickness $w_{\rm M} \approx 0.20$~m, and effective cavity factor $N_{\rm eff} \approx 300$~\cite{LIGOScientific:2014pky,Heisenberg:2023urf}.

A phenomenological dispersive contribution is parameterized by an effective photon mass term,
\begin{equation}
\mathcal{L}_{\rm mass} \supset \frac{1}{2}\, m_0^2 \,\mathcal{S}(\phi)\, A_\mu A^\mu ,
\end{equation}
which modifies the local dispersion relation and induces a frequency-dependent phase shift:
\begin{align}
    \Delta\varphi_{m_0}(t) &= \frac{m_0^2 N_{\rm eff}}{2\omega} \left[ \mathcal{I}_x(t) - \mathcal{I}_y(t) \right], \nonumber \\
    \mathcal{I}_{j}(t) &= \int_0^L \dd \ell \, \mathcal{S}\big( u(\ell \hat{\bm{j}}, t) \big),
\label{eq:phase_dispersive}
\end{align}
where $\omega=2\pi c/\lambda$.

Spatial gradients of the interaction energy density exert a dynamical force on the test masses~\cite{Grote:2019uvn}:
\begin{equation}
a_j(t)\simeq -c^2 g_{\rm DW}(\hat{\bm j}\cdot\hat{\bm n})\frac{d\mathcal{S}}{du}\Big|_{u=u(\bm r_j,t)}
\label{eq:acceleration}
\end{equation}
with $\bm r_x=L\hat{\bm x}$ and $\bm r_y=L\hat{\bm y}$.
The induced displacements satisfy $\ddot x_j=a_j$, and the differential arm length change is $\Delta L_{\rm CM}(t)=x_y(t)-x_x(t)$, giving $\Delta\varphi_{\rm CM}(t)=(4\pi/\lambda)\Delta L_{\rm CM}(t)$.

The total differential phase response is
\begin{equation}
    \Delta\varphi(t)=\Delta\varphi_{\rm size}(t)+\Delta\varphi_{m_0}(t)+\Delta\varphi_{\rm CM}(t),
\end{equation}
and the corresponding strain is $h(t)=(\lambda/4\pi L)\Delta\varphi(t)$.

\textcolor{black}{The planar-wall template assumes that the wall curvature is negligible across the detector network. For a curvature radius $R$ the deviation from planarity over the H1--L1 baseline $|\bm d|\approx3000$~km is $|\bm d|^{2}/(2R)$, which is small compared with the wall thickness $\delta_w\simeq\hbar c/m_\phi\approx2\times10^{2}$~km provided $R\gtrsim2\times10^{4}$~km. This is satisfied by any astrophysical wall, so the inferred parameters are insensitive to the global geometry of the wall network.}

More general quadratic interactions can also include couplings to fermion masses, $\sum_i \lambda_i\,\mathcal{S}(\phi)\,m_i \bar{\psi}_i\psi_i$. In this work, we restrict to the photon sector interaction and use $g_{\rm DW}$ as a single effective amplitude. A joint analysis that simultaneously fits multiple independent Standard Model couplings is left for future work. While the TDM posterior is sharply peaked in the mass scale parameters $m_\phi$ and $m_0$, it remains comparatively broad and only weakly constrained in $g_{\rm DW}$, so introducing fermion mass couplings $\lambda_i$ would mainly add further amplitude degeneracies and is unlikely to change the Bayes factor.

\section{Data conditioning and likelihood}

Data conditioning is adapted for the TDM analysis bandwidth using the publicly released noise power spectral density estimates and strain data~\cite{LIGOScientific:2025snk}. The data is downsampled to $400$~Hz, whitened using the estimated noise power spectral density, and bandpassed between $10$~Hz and $200$~Hz. The analysis is conducted on a $1.6$-second window around the respective trigger times, with a Tukey taper applied to the residuals to mitigate spectral leakage.

Our statistical framework assumes a Gaussian, unit variance noise model for the whitened time-domain samples. The signal in Livingston is coherently related to that in Hanford by a fixed rotation $R$ between the two detector frames and by the kinematic delay $\Delta t=(\bm d\cdot\hat{\bm n}_{\rm H1})/v_{\rm DW}$. The joint log-likelihood is
\begin{equation}
    \ln \mathcal{L}(d|\boldsymbol{\theta}) = -\frac{1}{2} \sum_{i \in \{\text{H1, L1}\}} \langle d_i - h_i(\boldsymbol{\theta}) | d_i - h_i(\boldsymbol{\theta}) \rangle,
\end{equation}
where the inner product denotes the sum of squares of the whitened time-domain residuals.

\section{Prior choices}

We employ an 8-parameter model for the TDM signal. The prior distributions are summarized in Table~\ref{tab:priors}.

Domain walls are extended planar objects that, unlike particle dark matter, are not captured by the galactic halo. Their velocity is set by formation dynamics and cosmological Hubble damping, which conserves $\gamma v\,a^3$ for a planar wall~\cite{Avelino:2014xda}, giving $v_{\rm DW}\propto(1+z_{\rm PT})^{-3}$. Since the symmetry-breaking scale $f$ that determines $z_{\rm PT}$ cancels in the detector response, $v_{\rm DW}$ cannot be independently constrained. We adopt a log-uniform prior on $v_{\rm DW}\in[10^{-4},\,10^{-1}]\,c$, spanning the range that produces signals in the LIGO band.

We adopt log-uniform priors for $m_\phi$, $m_0$, and $g_{\rm DW}$.
The scalar mass $m_\phi$ sets the wall thickness $\delta_w\sim \hbar/(m_\phi c)$ and crossing time $\tau_w\sim \delta_w/v_{\rm DW}$, so we choose $\log_{10}(m_\phi/{\rm eV})\in[-13,-10]$ to cover transient durations relevant to our analysis window.
The photon sector parameter $m_0$ controls the in-wall effective photon mass; radio observations down to MHz energies imply that $m_0\gtrsim{\rm few}\times{\rm neV}$ would cause strong reflection, so we truncate at $\log_{10}(m_0/{\rm eV})\le -9$~\cite{Jaeckel:2016jlh}.
In our model $\delta\alpha/\alpha\simeq g_{\rm DW}\mathcal{S}(\phi)$ for $|g_{\rm DW}|\ll1$, hence $|g_{\rm DW}|\sim|\delta\alpha/\alpha|_{\max}$; we therefore take a conservative prior $\log_{10}g_{\rm DW}\in[-30,-24]$, consistent with existing null searches for transient $\alpha$-variations~\cite{Roberts:2019sfo}.

We assume an isotropic prior for the wall normal $\hat{\bm n}$, sampling $\cos\theta \sim {\rm Unif}[-1,1]$ and $\phi \sim {\rm Unif}[0,2\pi]$.
Requiring the interaction to vanish in the vacuum values of the domain wall potential implies
$\mathcal{S}(\phi_k)=0$ for $\phi_k=2\pi f\,k/N_\phi$, hence $2N_A/N_\phi\in\mathbb{Z}$, i.e. $N_{\rm ratio}\equiv N_A/N_\phi\in \tfrac12\mathbb{Z}$.

\begin{center}
\begin{tabular}{l l}
\hline\hline
\textbf{Parameter} & \textbf{Prior} \\
\hline
$v_{\rm DW}$ & LogUnif$[10^{-4},\,10^{-1}]\,c$\\
$\log_{10}(m_\phi/\text{eV})$ & $[-13, -10]$ \\ 
$\log_{10}(m_0/\text{eV})$ & $[-12, -9]$ \\
$N_{\rm ratio}$ & $\left\{\frac{3}{2}, 2, \frac{5}{2}, 3, \frac{7}{2} \right\}$ \\
$\log_{10}(g_{\rm DW})$ & $[-30, -24]$ \\
$(\cos\theta,\phi)$ & $\cos\theta\sim{\rm Unif}[-1,1],\ \phi\sim{\rm Unif}[0,2\pi]$\\
$t_0$ & ${\rm Unif}[-0.8,0.8]~\mathrm{s}$ \\
\hline\hline
\end{tabular}
\captionof{table}{Priors for the TDM analysis.}
\label{tab:priors}
\end{center}

For the BBH hypothesis, we use the NRSur7dq4 waveform model~\cite{Varma:2019csw}. 
The LIGO-Virgo-KAGRA (LVK) Collaboration analysis of GW231123 considers five different BBH waveform models and also performs injection studies using NRSur7dq4 signals~\cite{LIGOScientific:2025rsn}. 
While different BBH approximants can lead to small shifts in inferred parameters, such differences are expected to be negligible for our purposes compared to the difference between the BBH and domain wall hypotheses. The BBH signal is described by a 15-parameter model, including masses, spins, and extrinsic parameters. We adopt the standard LVK BBH parameter estimation priors, uniform in spin magnitudes and redshifted component masses (with chirp mass prior extended to $250$ $M_{\odot}$), isotropic in spin orientations, sky position, and binary orientation, and with a luminosity distance prior uniform in comoving volume using a flat $\Lambda$CDM cosmology~\cite{LIGOScientific:2021usb,KAGRA:2021vkt}.

For GW190521, we follow the same procedure using publicly available O3a strain data~\cite{LIGOScientific:2020iuh}. The analysis window is $1.6$~s centered on the trigger time. Prior ranges are identical to those used for GW231123. For the joint analysis reported in Table~I of the main text, the scalar masses $m_\phi$, $m_0$, and coupling $g_{\rm DW}$ are shared across events, while the wall velocity $v_{\rm DW}$, arrival time $t_0$, and orientation $(\cos\theta,\phi)$ are sampled independently for each event. We perform several robustness checks on these results. Extending the lower bound of the $g_{\rm DW}$ prior to $\log_{10}g_{\rm DW}=-31$ does not shift the posterior or alter the Bayes factor. We also verify parameter recovery by injecting a simulated TDM signal at the maximum likelihood parameters into off-source noise and confirming that the pipeline recovers the injected values within the posterior credible intervals.

A comparison with~\cite{Heisenberg:2023urf}, which analysed GW190521 in a broader 32-s window at 4096-Hz sampling, shows good qualitative agreement while exhibiting expected quantitative differences. That work found $\log_{10}\mathcal{B}_{\rm BBH/TDM}\approx 9.1$ for GW190521, compared with $11.3$ in the present analysis. The quantitative shift is a direct consequence of the different analysis choices: the shorter 1.6-s window used here is adapted to the domain-wall crossing timescale and includes less off-signal data, altering the effective number of degrees of freedom available to each model. Crucially, both analyses agree on the qualitative pattern: the per-event evidence favours BBH, yet the preference is atypically weak relative to what matched BBH injections would predict. The best-fit discrete parameter also shifts: Ref.~\cite{Heisenberg:2023urf} found $N_{\rm ratio}=2.5$, whereas the present analysis assigns $85\%$ posterior weight to $N_{\rm ratio}=3$. Given the five-point discrete prior and the different data conditioning, these values are mutually consistent; $N_{\rm ratio}=2.5$ and $N_{\rm ratio}=3$ are adjacent prior points, and the shift does not affect the Bayes factor or the qualitative conclusions.

\section{Posterior corner plots}

As a pipeline validation, we perform a BBH parameter estimation for GW231123 using the same data conditioning, likelihood, and nested sampling configuration employed for the TDM analysis. The recovered parameters are consistent with the published LVK analysis~\cite{LIGOScientific:2025rsn}. Having validated the pipeline on the BBH hypothesis, we apply the same framework to both events for the TDM--BBH model comparison reported in Table~I of the main text.

\begin{figure}[!ht]
  \centering
  \includegraphics[width=\linewidth]{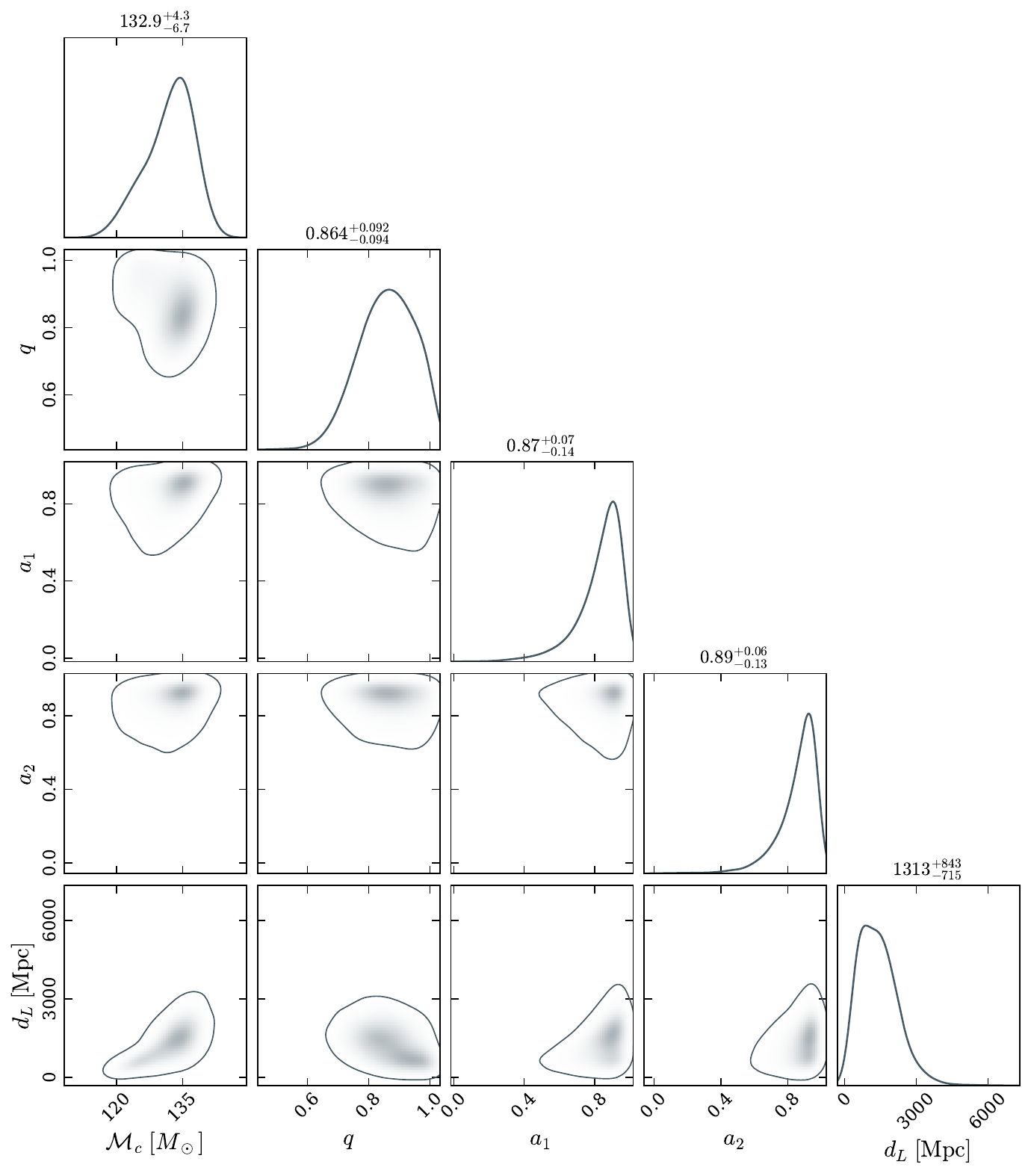}
  \caption{BBH posterior for GW231123 using the NRSur7dq4 waveform model, showing the source frame chirp mass $\mathcal{M}_c$, the mass ratio $q$, the dimensionless spin magnitudes $a_1$ and $a_2$, and the luminosity distance $d_L$.}
  \label{fig:bbh_corner}
\end{figure}

\begin{figure}[!ht]
  \centering
  \includegraphics[width=\linewidth]{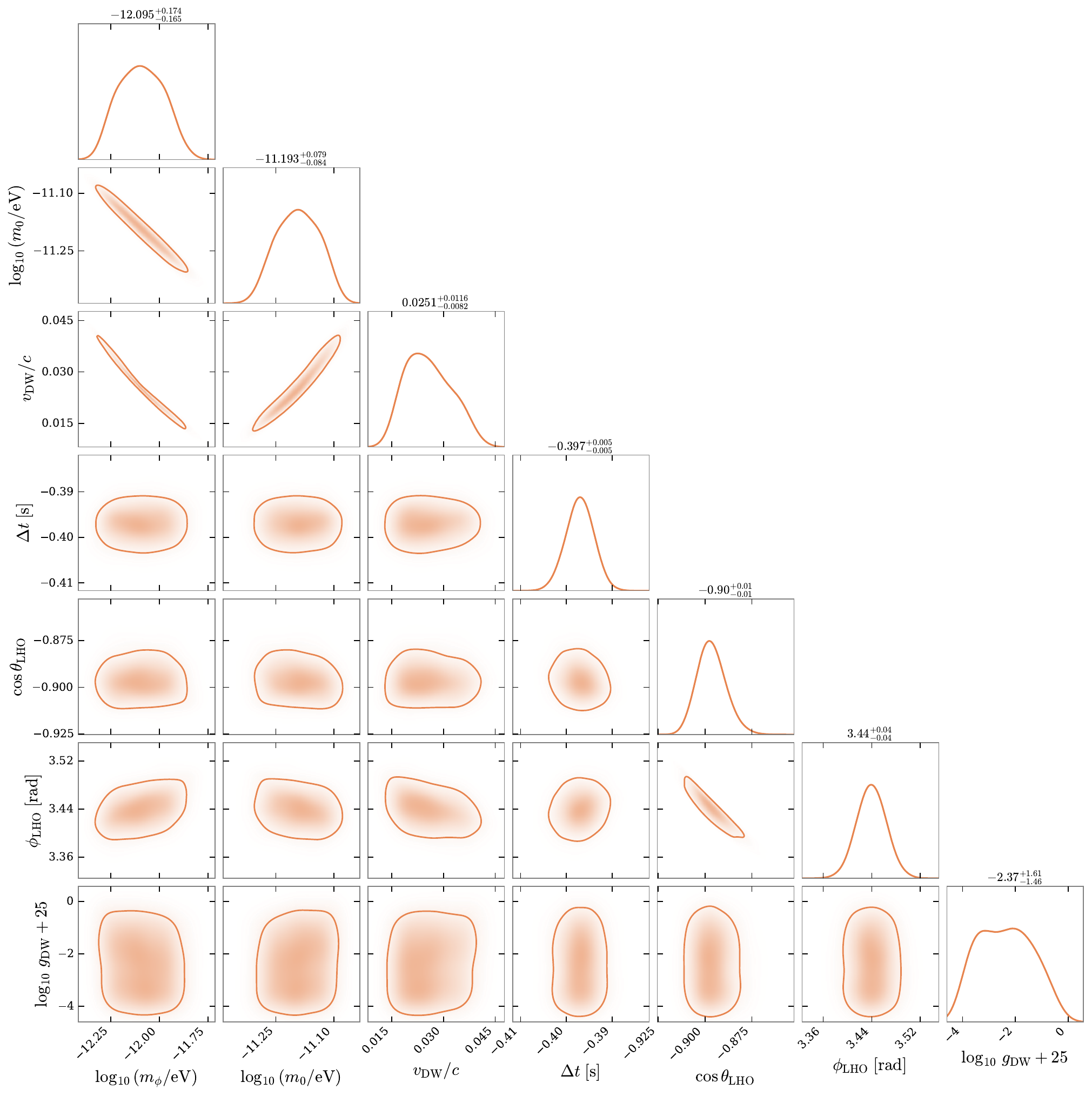}\\[4pt]
  \includegraphics[width=\linewidth]{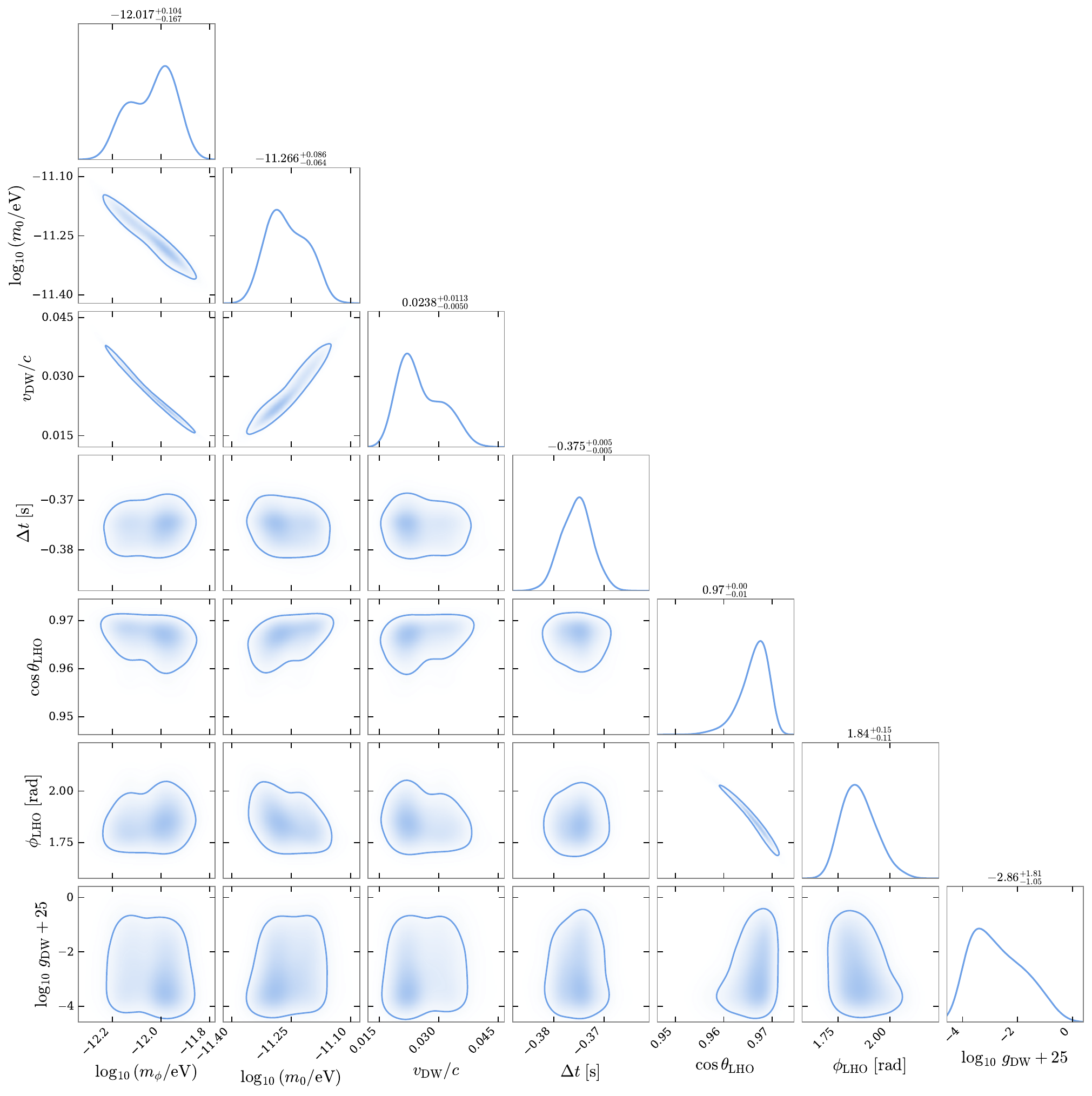}
  \caption{Posterior corner plots for the domain wall hypothesis for GW231123 (top) and GW190521 (bottom), showing 7 of the 8 sampled parameters: the scalar mass $m_\phi$, the photon sector parameter $m_0$, the wall speed $v_{\rm DW}/c$, the Hanford arrival time $t_0$, the wall normal angles $(\cos\theta,\phi)$ defining $\hat{\bm n}$ in the Hanford detector frame, and the effective coupling amplitude $g_{\rm DW}$. The discrete ratio $N_{\rm ratio}\equiv N_A/N_\phi$ concentrates entirely at $N_{\rm ratio}=3$ for GW231123 and assigns $85\%$ posterior weight to it for GW190521.}
  \label{fig:dw_corner}
\end{figure}
\section{Injection and recovery studies}

We use injection and recovery studies to calibrate the TDM versus BBH model comparison at the signal-to-noise ratios relevant for GW231123 and GW190521. For each event, we draw 50 noise-only off-source segments from the 192 s data stretch surrounding the trigger, excluding the on-source window. Into each segment, we inject the maximum a posteriori BBH waveform obtained from the corresponding BBH parameter estimation run. The injected data are conditioned in the same way as the real event and recovered under the TDM hypothesis. This ensemble gives the expected scatter of the recovered TDM parameters and of the Bayes factor when the underlying signal is a matched BBH waveform.

\begin{figure}[!ht]
  \centering
  \includegraphics[width=\linewidth]{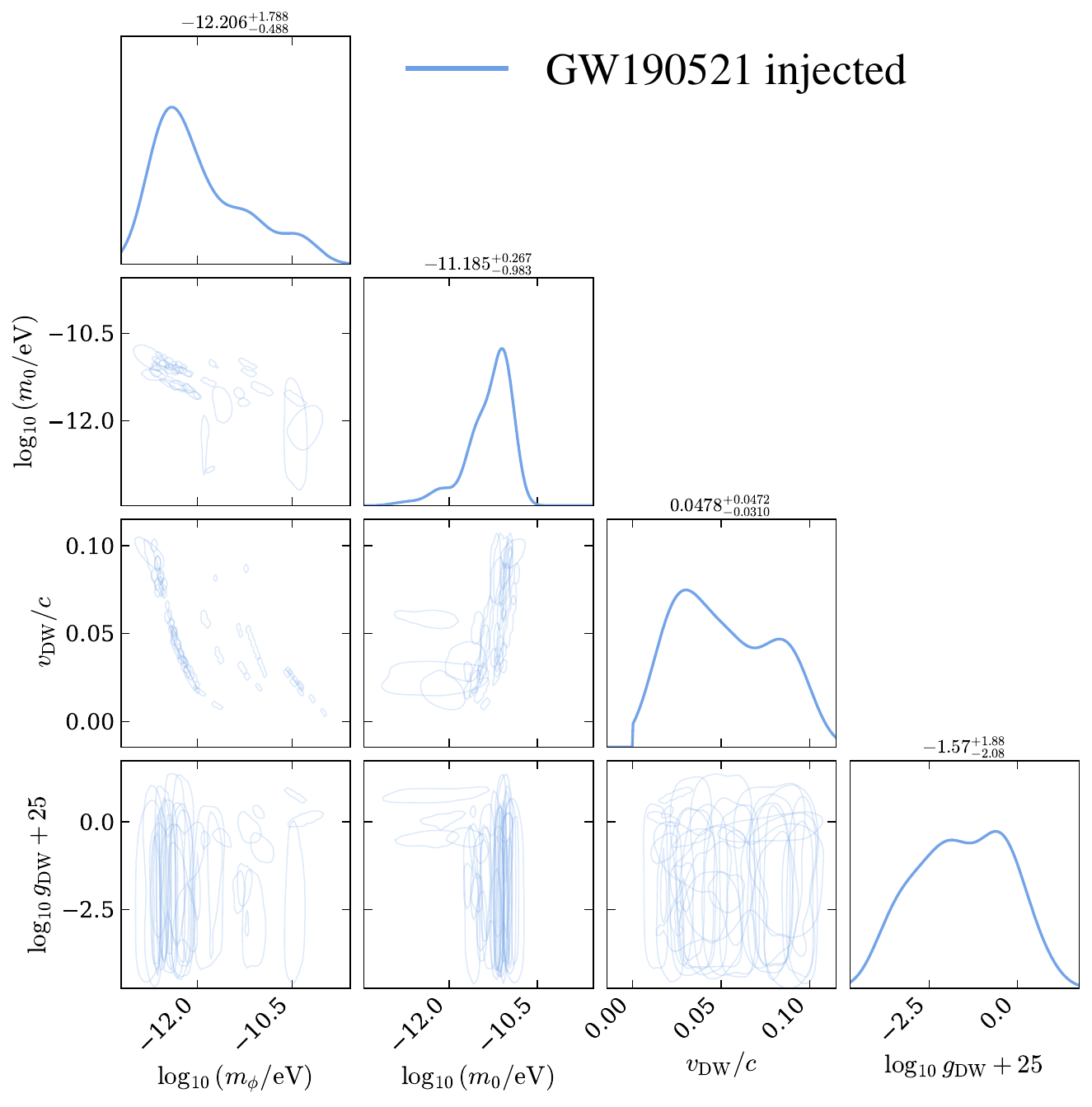}
  \includegraphics[width=\linewidth]{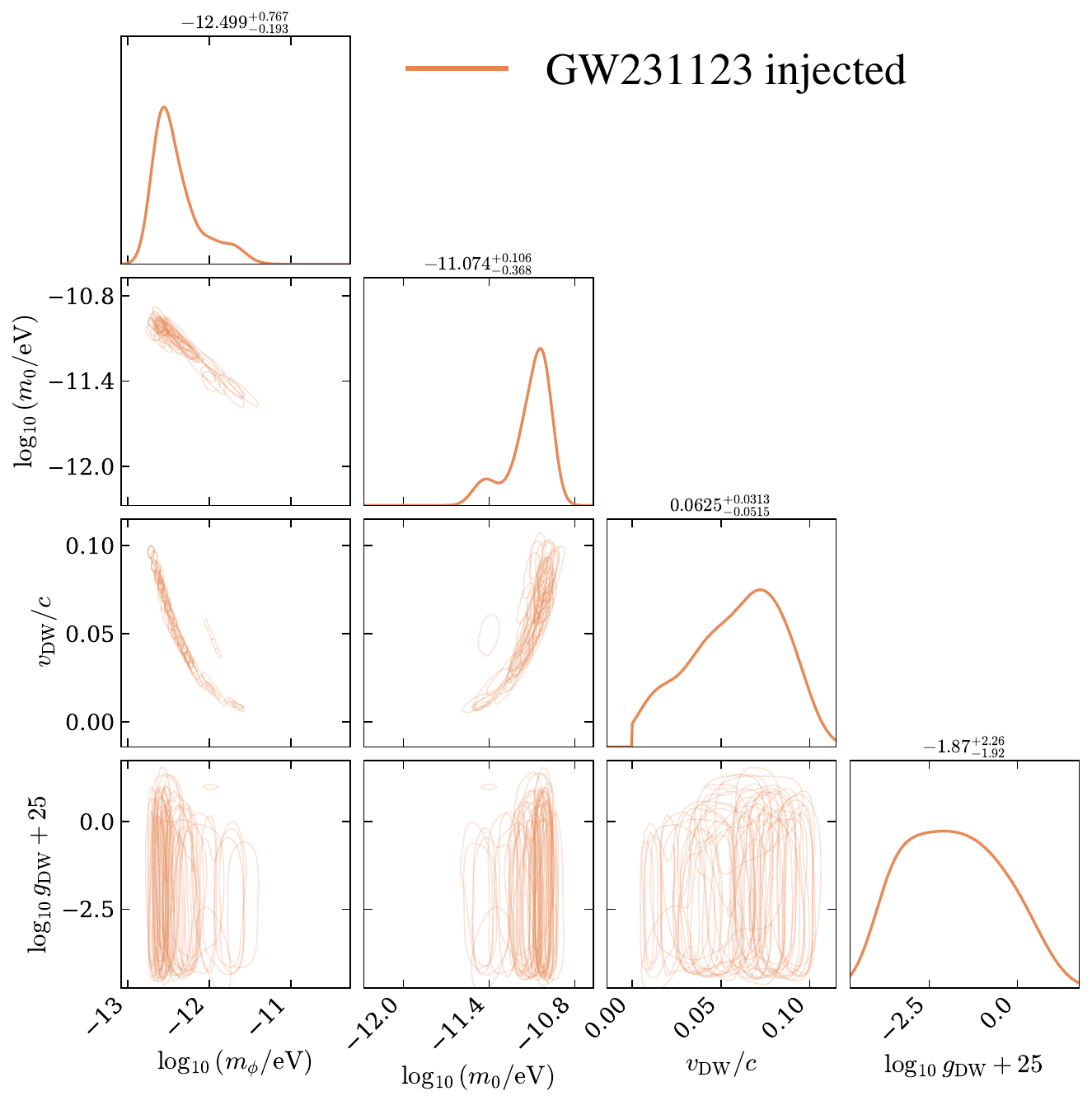}
  \caption{TDM recoveries of BBH-injected datasets for GW190521 and GW231123. Each panel shows 50 independent noise realizations. Off-diagonal contours enclose the $90\%$ credible region for individual realizations, while the diagonal panels show the marginalized one-dimensional distributions obtained by combining all realizations. The purpose of this test is to determine how the TDM parameter recovery fluctuates when the injected transient is drawn from the BBH hypothesis. The plotted parameters are the two domain-wall mass scales, the wall speed, and the effective coupling amplitude.}
  \label{fig:bbh_injected_tdm_recovery}
\end{figure}

The first test asks whether a matched BBH signal can produce TDM posteriors similar to those obtained from the real events. Figure~\ref{fig:bbh_injected_tdm_recovery} shows the TDM recoveries for the BBH-injected realizations. The spread of the contours shows how much the inferred TDM parameters fluctuate under independent nearby noise realizations when the injected signal is not a domain wall. This provides the reference distribution against which the observed TDM posteriors and Bayes factors should be interpreted.

As a complementary cross-recovery test, we inject the best-fit TDM waveform inferred from GW231123 into independent off-source noise segments and recover the resulting data under both the TDM and BBH hypotheses. This test is not used as a population model for the observed events. Its purpose is to check whether the pipeline can recover a clean TDM-like transient, and to identify where such a transient is mapped if it is forced into the BBH hypothesis.

\begin{figure}[!ht]
  \centering
  \includegraphics[width=\linewidth]{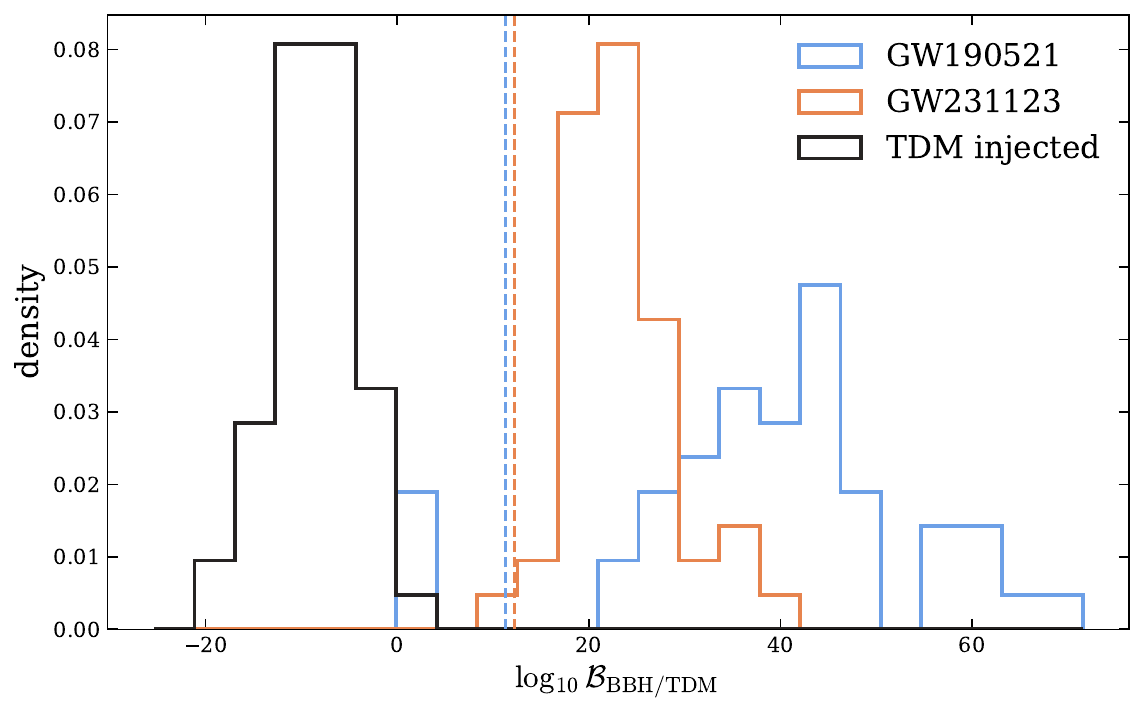}
    \caption{
    Distributions of $\log_{10}\mathcal{B}_{\rm BBH/TDM}$ from the injection
    campaigns. The GW190521 and GW231123 histograms show the BBH-injected
    ensembles; dashed vertical lines mark the observed-event values reported in
    Table~I of the main text. The observed values remain positive, so the
    single-event evidence favors the BBH hypothesis. However, they lie below the
    bulk of the BBH-injected distributions: $46/50$ GW190521 injections and
    $49/50$ GW231123 injections give Bayes factors at least as large as the
    observed values. The TDM-injected ensemble is concentrated at negative
    $\log_{10}\mathcal{B}_{\rm BBH/TDM}$, demonstrating that clean TDM injections
    are recoverable as TDM-favored signals. Thus, the observed events are favored
    by the BBH model, but not with the typical strength expected from the
    corresponding BBH-injected ensembles.
    }
  \label{fig:injection_bf}
\end{figure}

For each realization, we also compute
\begin{equation}
    \log_{10}\mathcal{B}_{\rm BBH/TDM}
    =
    \frac{\log Z_{\rm BBH}-\log Z_{\rm TDM}}{\ln 10}.
\end{equation}
Figure~\ref{fig:injection_bf} shows the resulting Bayes-factor distributions. Larger values favor the BBH hypothesis over the TDM hypothesis. For the BBH-injected ensembles, we quote the empirical injection percentile
\begin{equation}
    F_{\rm inj}(\mathcal{B}_{\rm obs}) =
    \frac{
    N\left(\log_{10}\mathcal{B}_{\rm BBH/TDM}
    \le
    \log_{10}\mathcal{B}_{\rm obs}\right)
    }{N_{\rm inj}} .
\end{equation}
This quantity measures how often a matched BBH injection gives a BBH preference no stronger than the one observed in the real data. It should not be interpreted as a detection significance for the TDM hypothesis.

\begin{figure}[!ht]
  \centering
  \includegraphics[width=\linewidth]{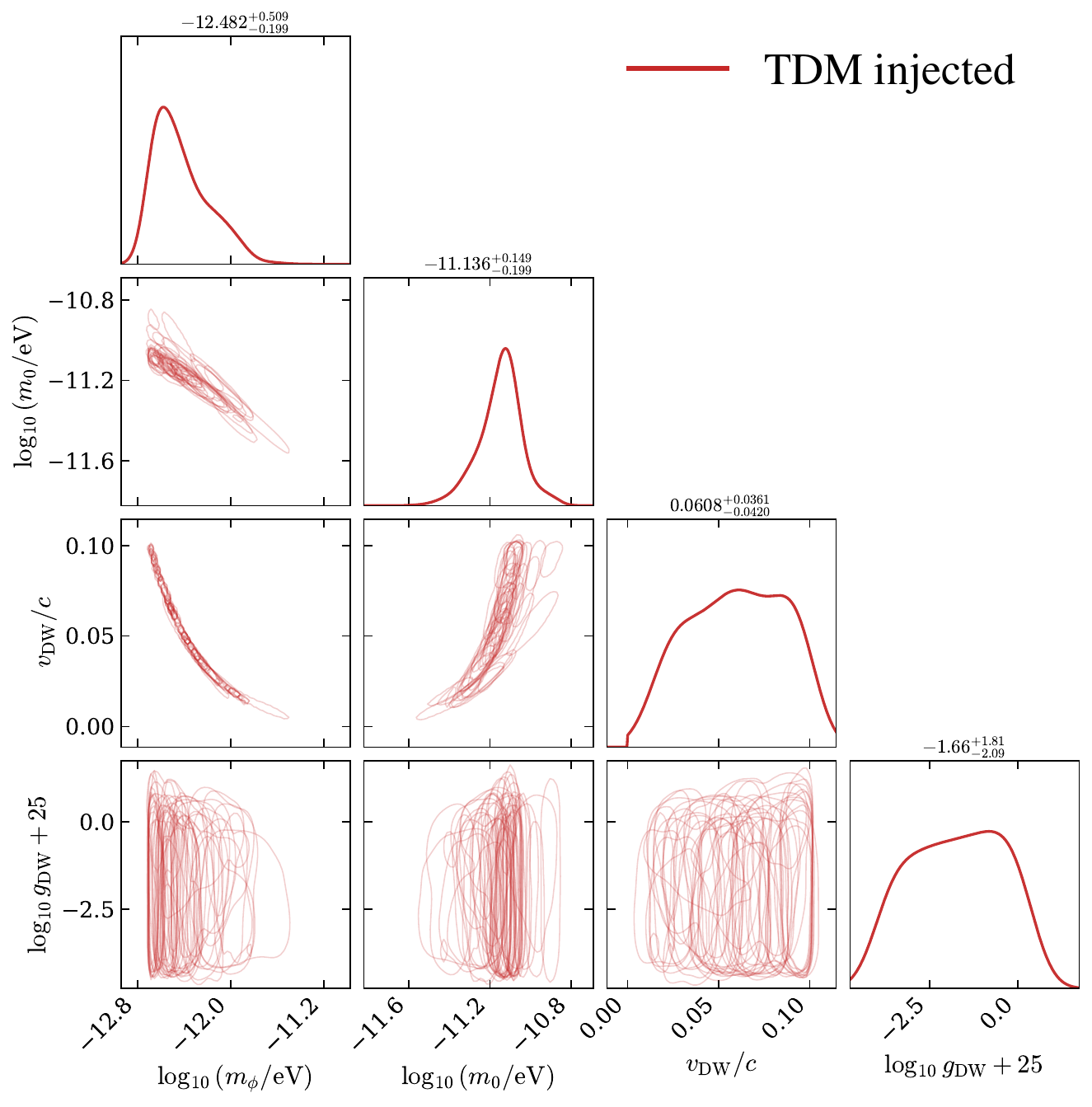}
  \includegraphics[width=\linewidth]{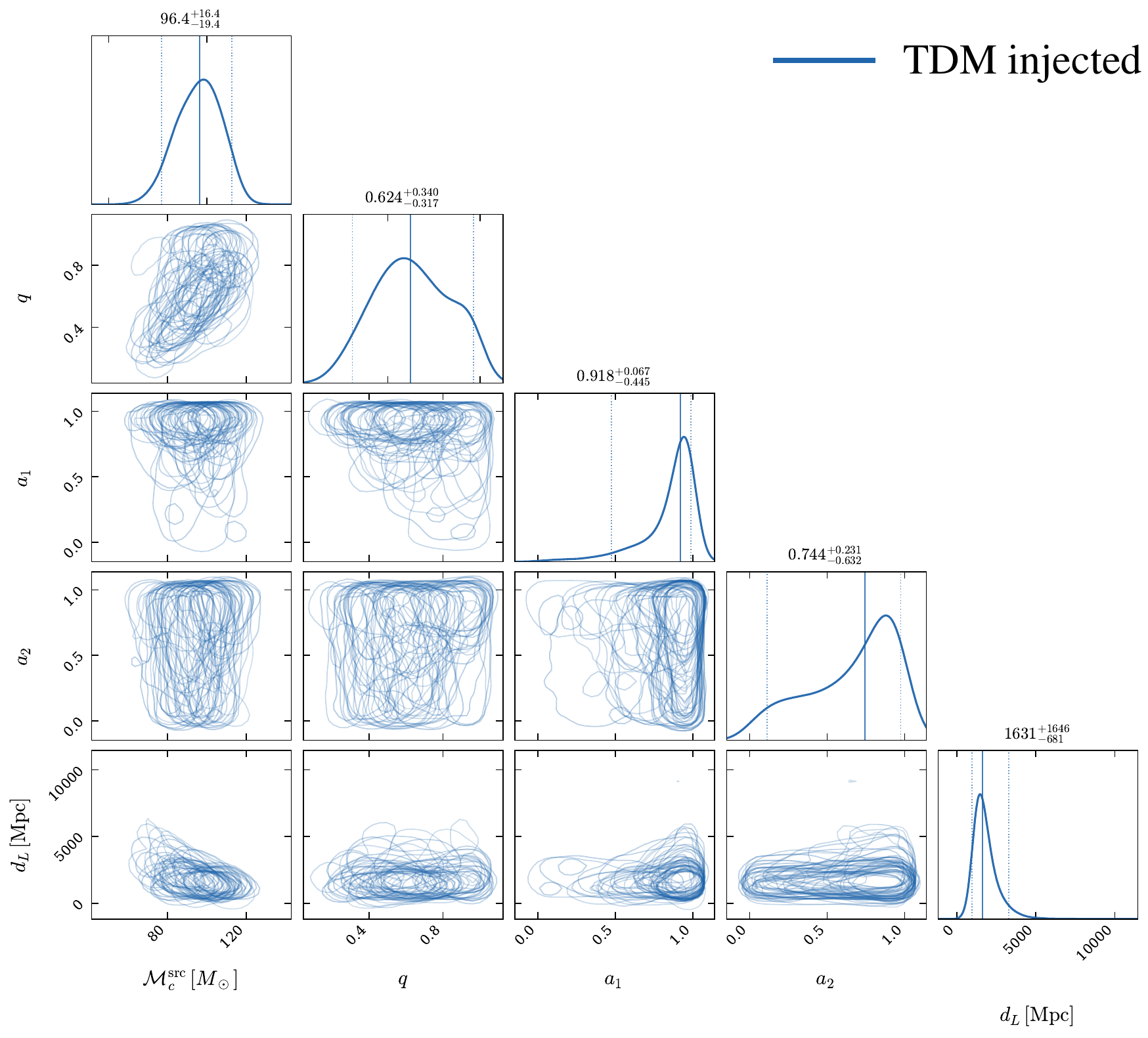}
    \caption{
    Recovery of TDM-injected datasets under the TDM hypothesis and under the BBH
    hypothesis. Each panel shows 50 independent noise realizations. The TDM
    recovery remains localized in the intrinsic domain-wall parameter space,
    showing that the analysis can recover the injected TDM region across
    independent noise realizations. The BBH recovery shows that the same TDM-like transient is mapped into a restricted apparent BBH region with broad mass-distance structure and support at large spin magnitudes. This is a morphological degeneracy of the BBH model for very short signals rather than an indication of physical black hole spin in the injected signal.
    }
  \label{fig:tdm_injected_recovery}
\end{figure}

For both real events, the Bayes factor is positive, so the per-event evidence 
prefers the BBH hypothesis over the TDM hypothesis. However, the observed 
values are not typical of the corresponding BBH-injected ensembles. For 
GW231123, the observed value $\log_{10}\mathcal{B}_{\rm BBH/TDM}=12.2$ lies 
below 49 of the 50 BBH-injected realizations, whose median is 
$\log_{10}\mathcal{B}_{\rm BBH/TDM}=23.1$. For GW190521, the observed value 
$\log_{10}\mathcal{B}_{\rm BBH/TDM}=11.3$ lies below 46 of the 50 BBH-injected 
realizations, whose median is $40.9$. Thus, the data favor the BBH hypothesis, 
but the strength of this preference is weaker than expected from matched maximum a posteriori BBH injections in nearby noise.

The TDM-injected realizations yield predominantly negative \(\log_{10}\mathcal{B}_{\rm BBH/TDM}\), with median \(-8.6\), showing that the analysis can identify a clean injected TDM transient when present. The real events are in an intermediate regime. Their per-event evidence values favor BBH, but the preference is weaker than expected from matched BBH injections in nearby noise. This motivates using the consistency of the shared TDM parameters across events as an additional diagnostic, rather than relying on a single-event Bayes factor alone.

\clearpage

\bibliographystyle{apsrev4-2}
\bibliography{bibliography}